\documentclass[lettersize,journal]{IEEEtran}
\usepackage{amsmath,amsfonts}
\usepackage{algorithmic}
\usepackage{algorithm}
\usepackage{array}
\usepackage{textcomp}
\usepackage{stfloats}
\usepackage{url}
\usepackage{verbatim}
\usepackage{graphicx}
\usepackage{cite}
\usepackage{tabularx}

\ifCLASSINFOpdf
\graphicspath{{./Figures/}}

\else

\fi

\usepackage{amsmath}
\usepackage{array}
\usepackage{ragged2e}

\newcolumntype{L}{>{\RaggedRight\arraybackslash}p{4cm}}
\newcolumntype{M}{>{\RaggedRight\arraybackslash}p{1.4cm}}
\newcolumntype{S}{>{\RaggedRight\arraybackslash}p{0.6cm}}

\usepackage{color,soul}
\usepackage{float}
\usepackage{booktabs}
\usepackage{multirow}
\usepackage{color}
\usepackage[usenames,dvipsnames,svgnames,table]{xcolor}
\usepackage{listings}
\usepackage{caption}
\usepackage{subcaption}
\usepackage[pagebackref=true,breaklinks=true,colorlinks,bookmarks=false]{hyperref}
\hypersetup{
    colorlinks=true,
    filecolor=magenta,      
    urlcolor=magenta,
}

\usepackage{hyperref}

\begin{document}

\title{CLARE: Cognitive Load Assessment in Real-time with Multimodal Data}

\author{Anubhav Bhatti, Prithila Angkan, Behnam Behinaein, Zunayed Mahmud, Dirk Rodenburg, Heather Braund, P. James Mclellan, Aaron Ruberto, Geoffery Harrison,  Daryl Wilson, Adam Szulewski, Dan Howes, Ali Etemad,~\IEEEmembership{Senior~Member,~IEEE}, Paul Hungler
\IEEEcompsocitemizethanks{
	\IEEEcompsocthanksitem A. Bhatti, B. Behinaein, Z. Mahmud, P. Angkan, and A. Etemad are with the Department of Electrical and Computer Engineering and Ingenuity Research Labs, Queen's University, Kingston, Ontario, Canada.\protect\\
	E-mails: {anubhav.bhatti, 9hbb, zunayed.mahmud, prithila.angkan, ali.etemad}@queensu.ca
	\IEEEcompsocthanksitem D. Rodenburg, J. Mclellan, P. Hungler are with Ingenuity Research Labs, Queen's University, Kingston, Ontario, Canada.\protect\\
	E-mails: {d.rodenburg, james.mclellan, paul.hungler}@queensu.ca
	\IEEEcompsocthanksitem   G. Harris,  D. Wilson is with the Department of Psychology, Queen's University, Kingston, Ontario, Canada.\protect\\
	E-mails: {8gh3@queensu.ca, daryl.wilson}@queensu.ca
	\IEEEcompsocthanksitem  H. Braund, A. Ruberto, A. Szulewski, D. Howes are with the School of Medicine, Queen's University, Kingston, Ontario, Canada.\protect\\
	E-mails: {heather.braund, a.ruberto, adam.szulewski, d.howes}@queensu.ca

}
}

\IEEEtitleabstractindextext{%
\begin{abstract}
 We present a novel multimodal dataset for Cognitive Load Assessment in REal-time (CLARE). The dataset contains physiological and gaze data from 24 participants with self-reported cognitive load scores as ground-truth labels. The dataset consists of four modalities, namely, Electrocardiography (ECG), Electrodermal Activity (EDA), Electroencephalogram (EEG), and Gaze tracking. To map diverse levels of mental load on participants during experiments, each participant completed four nine-minutes sessions on a computer-based operator performance and mental workload task (the MATB-II software) with varying levels of complexity in one minute segments. During the experiment, participants reported their cognitive load every 10 seconds. For the dataset, we also provide benchmark binary classification results with machine learning and deep learning models on two different evaluation schemes, namely, 10-fold and leave-one-subject-out (LOSO) cross-validation. Benchmark results show that for 10-fold evaluation, the convolutional neural network (CNN) based deep learning model achieves the best classification performance with ECG, EDA, and Gaze. In contrast, for LOSO, the best performance is achieved by the deep learning model with ECG, EDA, and EEG. 
\end{abstract}

}

\maketitle

\IEEEdisplaynontitleabstractindextext

\IEEEpeerreviewmaketitle

\section{Introduction}
\IEEEPARstart{C}{ognitive load} assessment is an essential aspect of affective computing, defined as the amount of mental effort utilized in the working memory during performance of a task.
Real-time assessment of cognitive load can enhance human-machine interactions, for instance, in training systems, education, transportation, automation, robotics, aerospace, etc. \cite{mckendrick2019theories, fridman2018cognitive, ahmad2020framework}. 
For example, quantifying the learners' cognitive load in such training systems can lead to optimization and personalization of learning content, which in turn can lead to better and more efficient learning outcomes \cite{brunken2010measuring, teigen1994yerkes, sweller1994cognitive, szulewski2020theory}. 

For cognitive load assessment, several tasks have been used to introduce varying amounts of mental workload on participants, such as n-back tasks, visual cue tasks, games, arithmetic, multiple-choice tests, and reading exercises, \cite{gjoreski2020datasets, mijic2019mmod, szulewski2015use}. 
For measuring cognitive load, three general approaches are utilized: self-reporting using tools such as the National Aeronautics and Space Association Task Load Index (NASA TLX) \cite{hart1988development} or Subjective Cognitive Load Scores (PAAS) \cite{paas1992training}, secondary behavioural performance, and the analysis of a range of physiological signals \cite{paas2016cognitive}.

\textcolor{black}{Prior work has shown that cognitive load and affect are closely related and can influence each other \cite{plass2019four}. One recent study showed the transferability of learned representations between these two domains \cite{pulver2023eeg}, where emotion datasets were used for pre-training encoders which were then used for downstream cognitive load classification. This study indicated the close relationship between cognitive load and affect from a computational perspective. Similarly, from a physiological perspective, the link between cognitive load and affect has been well-established \cite{zhao2023augmented, li2021sensitivity, mitchell2007psychological, matuliauskaite2011analysis}. }
\textcolor{black}{
Mental workload alters the sympathetic nervous system activities, which results in changes in Electrocardiogram (ECG), Electrodermal Activity (EDA), also known as Galvanic Skin Response (GSR), gaze, pupil size, Electroencephalogram (EEG), and other physiological responses. These signals can be used to estimate the cognitive load and emotional states using machine learning and deep learning algorithms \cite{ross2019toward, antonenko2010using, perkhofer2019using, krejtz2018eye, bhatti2022attx, sarkar2022avcaffe} and can be easily monitored using off-the-shelf wearables.}

\textcolor{black}{Recent advances allow researchers to utilize various machine learning algorithms to classify a user’s cognitive workload using physiological measures. Despite the potential benefits and use cases of automated cognitive load measurement, the lack of annotated cognitive load datasets has been a roadblock towards creating machine learning systems capable of estimating cognitive load with high confidence. Recent studies use self-reported cognitive load scores as ground truth labels; however, these scores are generally recorded upon completion of the experiment, and are thus unable to capture changes in cognitive load throughout the data collection process. This approach can affect the fidelity of the labels since it is administered retrospectively, which can introduce inaccuracies due to lapses in memory and recency bias \cite{mckendrick2019theories}. Additionally, the lack of frequent labels captured during the performed activities inhibits the development of machine learning systems capable of \textit{real-time} estimation of cognitive load.
} 

To overcome these problems, we introduce a multimodal dataset on wearable-based \textbf{C}ognitive \textbf{L}oad \textbf{A}ssessment in \textbf{RE}altime (CLARE). In this dataset, we collected four different data modalities, i.e., ECG, EDA, EEG, and gaze data, from 24 participants while performing MATB-II tasks designed to induce varying amounts of cognitive load. During the experiments, each participant completed four 9-minute sessions, comprised of different cognitive tasks with varying mental effort requirements. We documented the participants' self-reported subjective cognitive scores on a 9-point Likert scale at 10-second intervals during each session. As a consequence, our dataset differs from the few cognitive load datasets that have been made publicly available. Our methodology collects subjective self-reported cognitive load scores at frequent intervals throughout the experiments, instead of the commonly used approach of retrospectively recording scores. Given its frequency, this approach better facilitates the development of real-time machine learning solutions for cognitive load assessment.

Our main contributions in this paper are as follows: 
\begin{enumerate}
    \item \textcolor{black}{We present a new multimodal dataset on cognitive load assessment. The dataset comprises ECG (512 Hz), EDA (128 Hz), EEG (256 Hz), and Gaze tracking (50 Hz) data from 24 participants with self-reported cognitive load scores at regular intervals of 10 seconds.} 

	\item \textcolor{black}{We analyze and provide insights to understand the distribution of reported ground truths with respect to the complexity of performed tasks. We provide machine learning and deep learning baselines for estimating cognitive load in uni-modal and multimodal setups using two evaluation schemes, namely, 10-fold and leave-one-subject-out cross-validation.}
	
	\item \textcolor{black}{We make the dataset publicly available to facilitate and promote real-time cognitive load estimation research and contribute to the field.}

\end{enumerate}

{\renewcommand{\arraystretch}{1.4}
\begin{table*}[!t]
    \centering
    \caption{Affect and cognitive load datasets}
    \begin{tabularx}{\textwidth}{p{0.15\textwidth}llXXX}
    \midrule[.0025\linewidth]
    \textbf{Dataset} & \textbf{Year} & \textbf{Subs} & \textbf{Affective States} & \textbf{Modalities} & \textbf{Stimuli} \\ 
    \midrule[.0025\linewidth]
    MAHNOB-HCI\cite{soleymani2011multimodal} & 2011 & 27 & Arousal, valence, dominance & EEG, EDA, RESP, ST, gaze, audio, video & Movie clips\\
    
    DEAP \cite{koelstra2011deap} & 2011 & 32 & Arousal, valence, like/dislike, dominance, familiarity, face video & EEG, EOG, EMG, GSR, RESP, BVP, ST & Music video\\
    
    SWELL-KW \cite{koldijk2014swell} & 2014 & 25 & Task load, mental effort, emotion, stress & ECG, EDA, computer logging, facial, postures & Preparing presentations, reports, email, information searching\\
    
    DECAF\cite{abadi2015decaf} & 2015 & 30 & Arousal, valence, dominance & ECG, EMG, EOG, MEG, IR face video & Music video \& videos\\
    
    DREAMER \cite{katsigiannis2017dreamer} & 2017 & 23 & Arousal, valence, dominance & EEG, ECG & Movie clips\\
    
    AMIGOS \cite{correa2018amigos} & 2018 & 40 & Arousal, valence, dominance, familiarity, liking, basic emotions & ECG, EDA, EEG, audio, visual, depth (Kinect) & Videos\\
    
    WESAD \cite{schmidt2018introducing} & 2018 & 15 & Neutral, stress, amusement & \textbf{Chest}: ECG, EDA, EMG, ST, ACC, RESP; \textbf{Wrist}:  EDA, BVP, ACC, ST & Trier social stress test\\
    
    MPED \cite{song2019mped} & 2019 & 23 & Joy, funny, anger, disgust, fear, sad, neutrality & ECG, EDA, EEG, RESP & Videos\\
    
    CASE \cite{sharma2019dataset} & 2019 & 30 & Arousal, Valence & ECG, EDA, BVP, EMG, RESP, ST & Videos\\ \midrule

    CLAS \cite{markova2019clas} & 2019 & 62 & Negative emotions, mental strain, high cognitive effort & ECG, EDA, PPG, ACC & Math \& logic problems, stroop test, images, audio-videos\\
    
    CogLoad \cite{gjoreski2020datasets} & 2020 & 23 & Cognitive load, personality traits & EDA, ACC, ST, RR & Various cognitive load tasks\\
    
    Snake \cite{gjoreski2020datasets} & 2020 & 23 & Cognitive load & EDA, ACC, ST, RR & Snake game\\
    
    Kalatzis et al. \cite{kalatzis2021database} & 2021 & 26 & Cognitive load & ECG, RR & MATB-II\\  

    \textcolor{black}{CL-Drive} \cite{angkan2024multimodal} & \textcolor{black}{2024} & \textcolor{black}{21} & \textcolor{black}{Cognitive load} & \textcolor{black}{ECG, EDA, EEG, Gaze} & \textcolor{black}{Driving simulator}\\  

    \midrule

    CLARE (Ours) & 2022 & 24 & Cognitive load & ECG, EDA, EEG, Gaze & MATB-II\\ \bottomrule[.0025\linewidth]
    \end{tabularx}
    \label{tab:datasets-review}
\end{table*}
}

\section{Related Work} \label{sec:related-work}
Open datasets are crucial for conducting research, reproducing and verifying results in scientific communities. As mentioned previously, human cognitive load and affective states are interconnected and influence each other. Therefore, this section presents an overview of publicly available datasets containing biological signals for emotion recognition and cognitive load classification. Table \ref{tab:datasets-review} summarizes all the public datasets on emotion recognition and cognitive load assessment.

\subsection{General Affect Datasets}
First, we give an overview of datasets that cover a range of affective states. MAHNOB-HCI \cite {soleymani2011multimodal}  is a multimodal database containing physiological signals (ECG, EDA, RESP, ST, and EEG), facial videos, audio signals, and gaze data from 27 participants. In this dataset, movie clips were used as a stimulus for triggering affective states in participants and self-reported values of arousal, valence, and dominance were collected. Classical machine learning algorithms provided baselines for uni-modal and multimodal classification of emotions. DEAP dataset \cite{koelstra2011deap}  comprises a collection of biological (EEG, EOG, EMG, EDA, RESP, BVP, and ST) and facial data from 32 participants. In this experiment, the participants watched music videos that were selected to stimulate emotions. The dataset includes a self-assessment of arousal, valence, like/dislike, dominance, and familiarity. They performed baseline uni-modal and multimodal classification of affect states using naive Bayes classifiers. Koldijk et al. introduced the SWELL-KW dataset \cite{koldijk2014swell} for investigating stress in a knowledge working environment by manipulating working conditions with stressors such as email interruptions and time pressure. SWELL-KW contains computer logging, facial expression, postures, and physiological signals (ECG and EDA) from 25 subjects. DECAF dataset \cite{abadi2015decaf} contains affect responses, such as valence, arousal, and dominance, of 30 subjects while watching music videos and video clips. This dataset contains ECG, EMG, EOG, magnetoencephalography (MEG), and infrared (IR) face videos. They also provide baseline scores for valance, arousal, and dominance classification using MEG signals.

A new wave of affective datasets were created and made public in 2017 and later. DREAMER \cite{katsigiannis2017dreamer} is a multimodal database of 23 subjects' affective states in response to movie clips. The physiological data, EEG and ECG, were captured during affect elicitation using portable, wearable, off-the-shelf wireless equipment. After each stimulus, the participants' self-assessments of their affective state valence, arousal, and dominance were recorded. They used classical machine learning to provide baselines for affect recognition using EEG and ECG. The AMIGOS dataset \cite{correa2018amigos}, is a multimodal dataset of affect, personality traits, and mood from 40 participants. In this dataset, affect states were elicited using short and long videos. Physiological signals (ECG, EDA, EEG), audio, and frontal HD video were recorded in this dataset. Both self-assessment of affective states (valence, arousal, dominance, familiarity, liking, and basic emotions) and external assessment of valence and arousal levels are provided in the dataset. 

WESAD \cite{schmidt2018introducing} multimodal dataset used wearable devices to collect physiological and motion data from 15 participants for stress detection. They collected data using chest-worn (ECG, EDA, EMG, ST, accelerometer (ACC), RESP) and wrist-worn (ACC, BVP, EDA, ST) sensors. The stress was elicited using Trier Social Stress Test \cite{kirschbaum1993trier} where the participants performed free speech and mental arithmetic operations in front of an audience. They recorded three different affective states, namely neutral, stress, and amusement. MPED \cite{song2019mped}, is a multimodal physiological emotion database of 23 participants. The dataset includes four physiological signals, ECG, EEG, EDA, and RESP. In this dataset, the participants watched videos that were chosen to elicit six discrete emotions. In CASE dataset \cite{sharma2019dataset} continuous annotation of emotions were collected from 30 participants while watching videos. The dataset included eight physiological signals (ECG, EDA, BVP, EMG, RESP, and ST) collected at a sampling rate of 1000Hz. The dataset used a joystick-based annotation method for real-time annotations of arousal and valence.

\subsection{Cognitive Load Datasets}
In this section, we review prior work that present datasets on cognitive load assessment. CLAS \cite{markova2019clas} is a multimodal dataset that collected physiological (ECG, EDA, PPG), and ACC data from 62 participants. During the experiment, to elicit emotions, mental strain, and high cognitive load, the participants performed various tasks such as solving mathematical and logical problems, taking the Stroop test, and watching images and audio-video clips. Gjoreski et al. \cite{gjoreski2020datasets} introduced two multimodal datasets, CogLoad and Snake, for facilitating research on cognitive load inference and personality traits. Both datasets collected physiological data (time interval between R-peaks (RR), EDA, ACC, and ST) from wrist-worn devices from 23 participants. In the CogLoad dataset, participants performed multiple tasks, such as 2- and 3-back tasks. In the Snake dataset, the participants played a game with varying levels of difficulty (easy, medium, and hard) on a smartphone. The participants' personality traits were recorded using questionnaires in both datasets, and their perceived cognitive load was collected using a self-reporting tool, the National Aeronautics and Space Association Task Load Index (NASA TLX) \cite{hart1988development}. Kalatzis et al. \cite{kalatzis2021database} introduced a database comprising data from 26 participants for cognitive load research. ECG and RR data were collected from the participants while they performed two tasks to induce low and high cognitive workloads using the MATB-II software. The participants' perceived cognitive load levels were recorded using the NASA-TLX. 
\textcolor{black}{CL-Drive \cite{angkan2024multimodal}, used four modalities EEG, ECG, EDA and Gaze data to classify cognitive load during driving in a vehicle simulator. The data was collected from 21 participants and the self-reported cognitive load scores were recorded every 10 seconds. The dataset contains 9 trials of 3 minuted each for each participant of varying cognitive load.}

\textcolor{black}{Overall, our review of the existing datasets on cognitive load reveals a number of limitations, which we address through CLARE in this paper. 
Specifically, we observe that some datasets such as \cite{kalatzis2021database} include a low number of modalities, whereas our dataset offers large number and diverse modalities for cognitive load classification. Additionally, most datasets in this area, for instance \cite{kalatzis2021database, markova2019clas, gjoreski2020datasets}, record cognitive load scores \textit{at the end of the experiment}, and do not provide labels that correspond the wearable signals throughout the experiments. This means that most existing datasets cannot be used to develop machine learning systems capable of \textit{real-time} classification of cognitive load. Our dataset, on the other hand, contains frequent ground-truth cognitive load scores recorded throughout the data collection process, which could enable the development of real-time systems.
}

\section{Experiment setup} \label{sec:experiment-setup}
\subsection{MATB-II Software}
\begin{figure*}[!t]
	\centering
	\includegraphics[width=.7\linewidth]{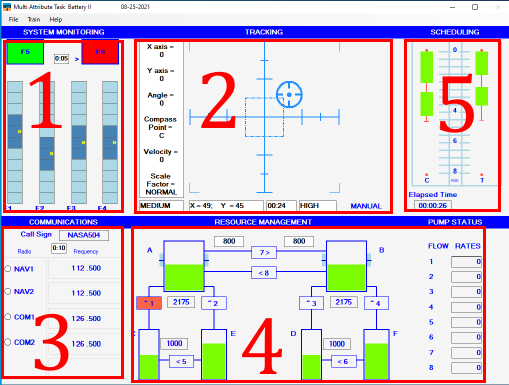}
	\caption{A snapshot of MATB-II interface. Region 1 shows the system monitoring task, region 2 depicts the tracking task, region 3 shows the communication task, region 4 shows the resource management task, and region 5 shows the scheduler part of the software.}
	\label{fig:matbgui}
\end{figure*}

To induce varying amounts of cognitive load, we used the Multi-Attribute Task Battery (MATB-II) tool \cite{santiago2011multi} developed by the National Aeronautics and Space Administration (NASA)\footnote{https://www.nasa.gov/ [Accessed: 2022-06-05]}. 
\textcolor{black}{MatB-II is a widely used software by researchers to induce cognitive load in participants. This tool was developed to facilitate research on human multi-task performance and mental workload and has been widely used for assessing emotions and cognitive load in various scientific disciplines \cite{bulikhov2023effect, kennedy2017making, chandra2015eeg, qu2021classification, yang2023mental}. MATB-II is able to induce different levels of cognitive load in participants by changing the complexity of the tasks performed. It consists of 5 different tasks which can be active individually or at the same time to control the complexity, making it ideal for our application.}
As shown in Figure \ref{fig:matbgui}, the MATB-II interface has five regions comprising four tasks and one scheduling region. These regions are explained as follows:

\begin{enumerate}
    \item{\textit{System monitoring task}: In this task, shown as region 1 in the Figure \ref{fig:matbgui}, there are four slider bars (F1 to F4), and two lights (F5 that turns grey/green denoting off/on and F6 that turns grey/red for on/off).
    The participant's task is to reset the sliding bars when the indicator reaches either end of the bar by clicking anywhere on the bar. 
    In secondary task, the participant needs to monitor and respond to the absence of green ``light'' (default state is ``on''), and the presence of red ``light'' (default state is ``off'') by clicking on the respective light boxes.} 
    \item{\textit{Tracking task}: In this task, shown in region 2 of Figure \ref{fig:matbgui}, there are two modes: automatic and manual. In the automatic mode, no attention is required from the participant, while in the manual mode, the participant needs to actively use the joystick to keep the crosshair inside the black square in the middle.}
    \item{\textit{Communication task}: In the communication task shown in region 3 of Figure \ref{fig:matbgui}, two types of audio cues play for the participants: relevant and irrelevant cues. The participant only needs to respond to the relevant cues starting with the call sign ``NASA504'' and ignore other call signs. Each call contains a communication channel (NAV1, NAV2, COM1, and COM2) and six digits for frequency. The participant must choose the channels for the relevant calls and subsequently enter the broadcast frequency in the frequency section.}
    \item{\textit{Resource management task}: In this task, shown in region 4 in Figure \ref{fig:matbgui}, the goal is to keep tanks A and B at the specified levels (marked with black bars on the side of tanks). These tanks keep depleting during the experiment and must be refilled using tanks C, D, E, and F. The participant can circulate the fuel in the tanks using pumps marked as 1-8. The arrow `$>$' shows the direction in which the pumps can circulate fuel. Tanks C and D have limited capacity but tanks E and F have unlimited capacity. The pumps can be toggled to `inactive' (idle shown with white colour) or `active' (pumping shown with green) states by clicking on them. The pumps can also break (red colour) during the experiment, making the pump unusable. After a specific time, the broken pumps are fixed automatically.}
    \item{\textit{Scheduling}: Region 5 shown in Figure \ref{fig:matbgui} shows the Scheduling section where the participants can see the incoming `communication' and `tracking' events as a function of a vertical timeline. Event onset is signaled by arrival at the top of the timeline. }
\end{enumerate}

The complexity of the experiment can be modulated by changing the number of occurrences of a task or the difficulty of the tasks. For instance, the frequency of turning the green and red lights can be changed in the system monitoring task. Also, the speed of the sliders on the bars can be changed. Several parameters can be modified in the tracking task, e.g., the number of times the tracking task switches from automatic mode to manual mode and the joystick's sensitivity and response speed. The number of relevant and irrelevant messages during an interval can be modified in the communication task. Similarly, for the resource management section, one can change the number of times the pump breaks.

\begin{figure}
	\centering
	\includegraphics[width=0.7\linewidth]{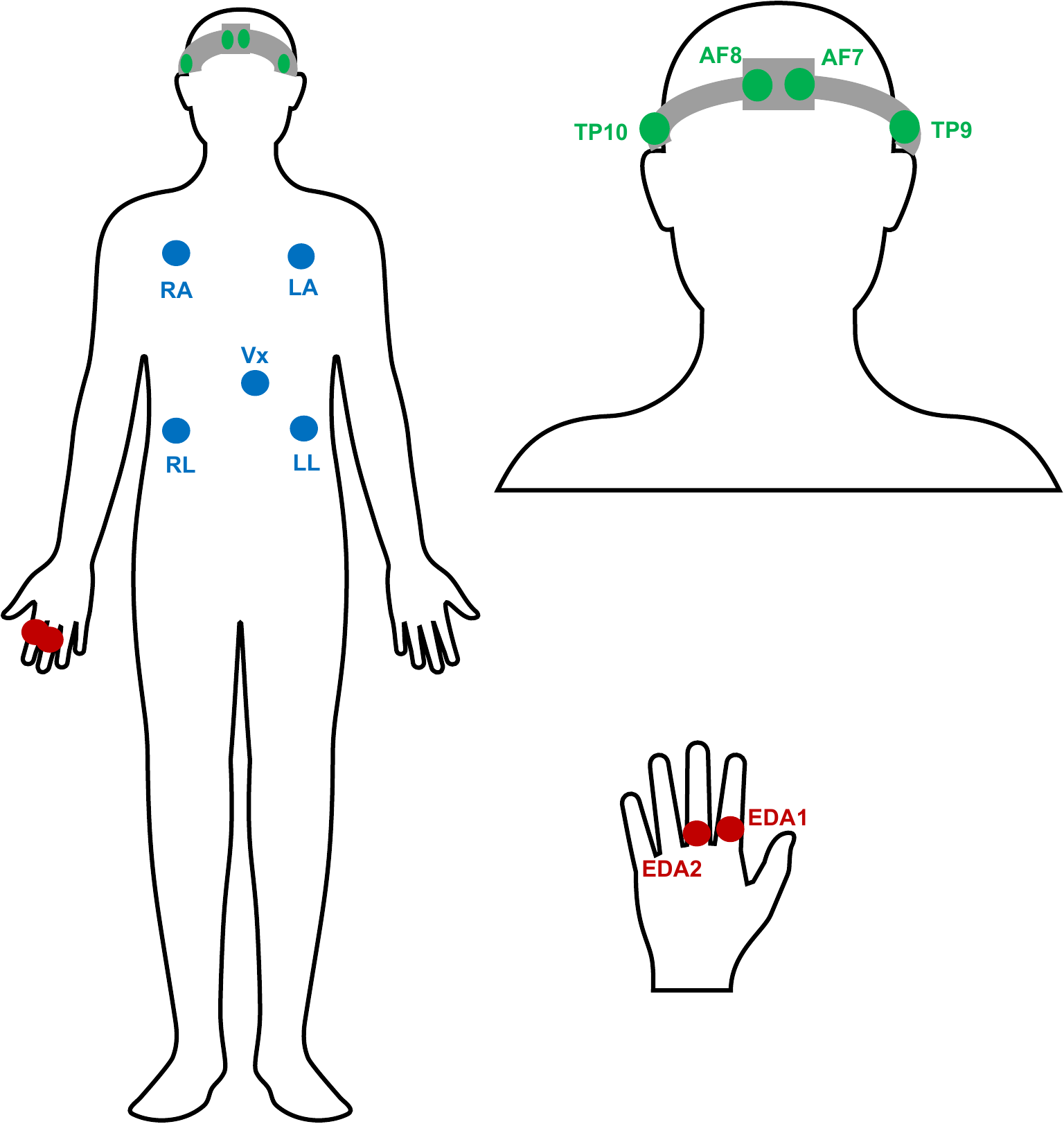}
	\caption{Sensor placements, ECG (in black), EDA (in red), and EEG (in green).}
	\label{fig:sensorsplacement}
\end{figure}

\begin{figure}[t]
    \centering
    \subfloat[Ascending complexity]
    {\includegraphics[width=0.45\linewidth]{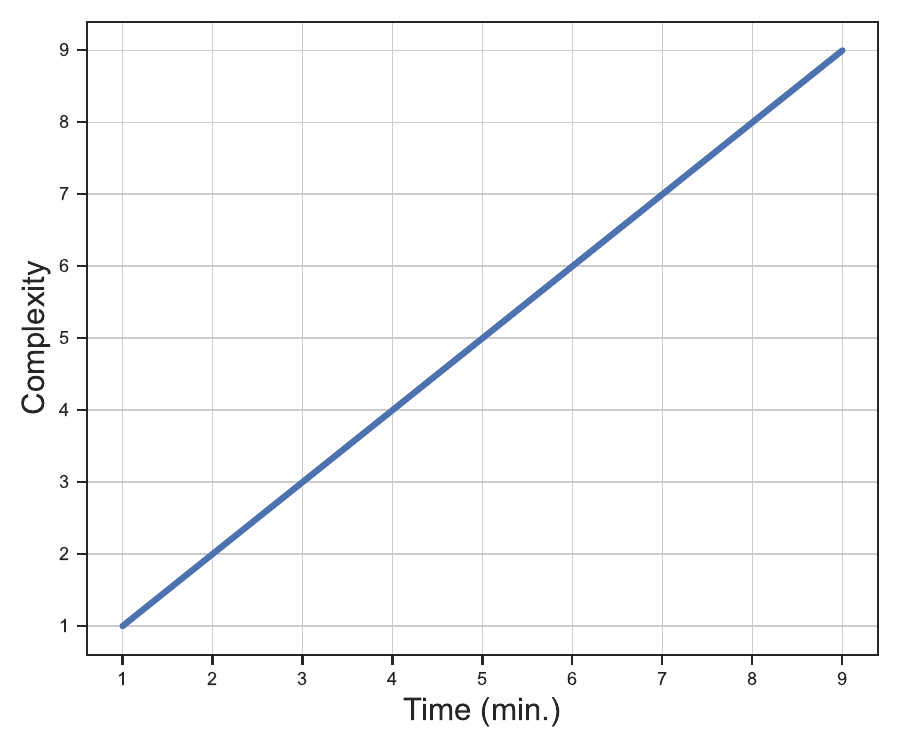}
    \label{fig:ascComplexity}}
    \subfloat[Sequence 1]
    {\includegraphics[width=0.45\linewidth]{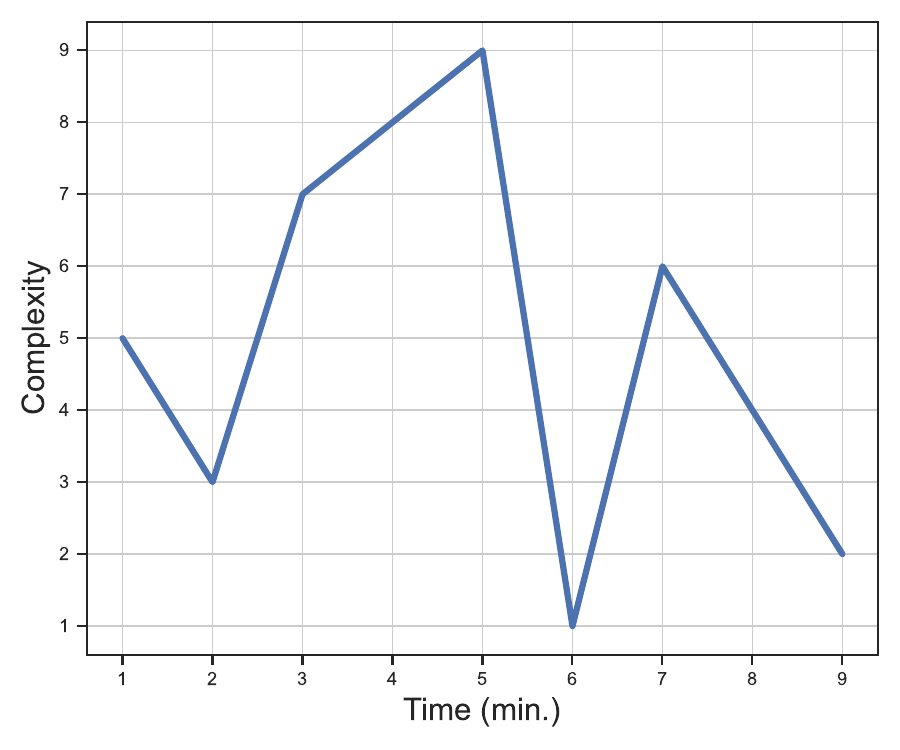}
    \label{fig:seq1}}

    \subfloat[Sequence 2]
    {\includegraphics[width=0.45\linewidth]{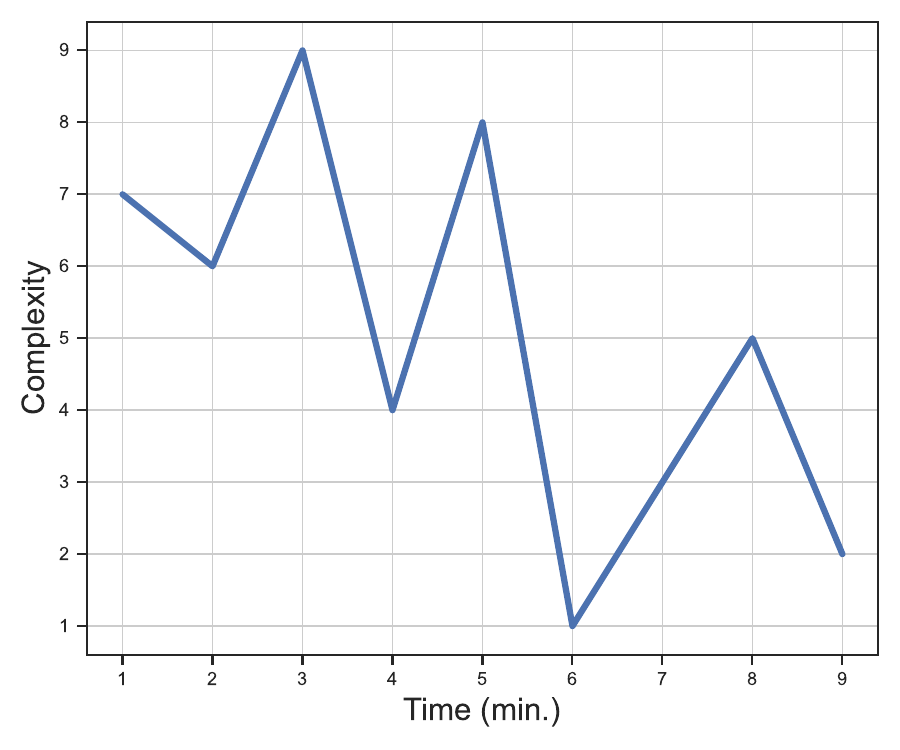}
    \label{fig:seq2}}
    \subfloat[Sequence 3]
    {\includegraphics[width=0.45\linewidth]{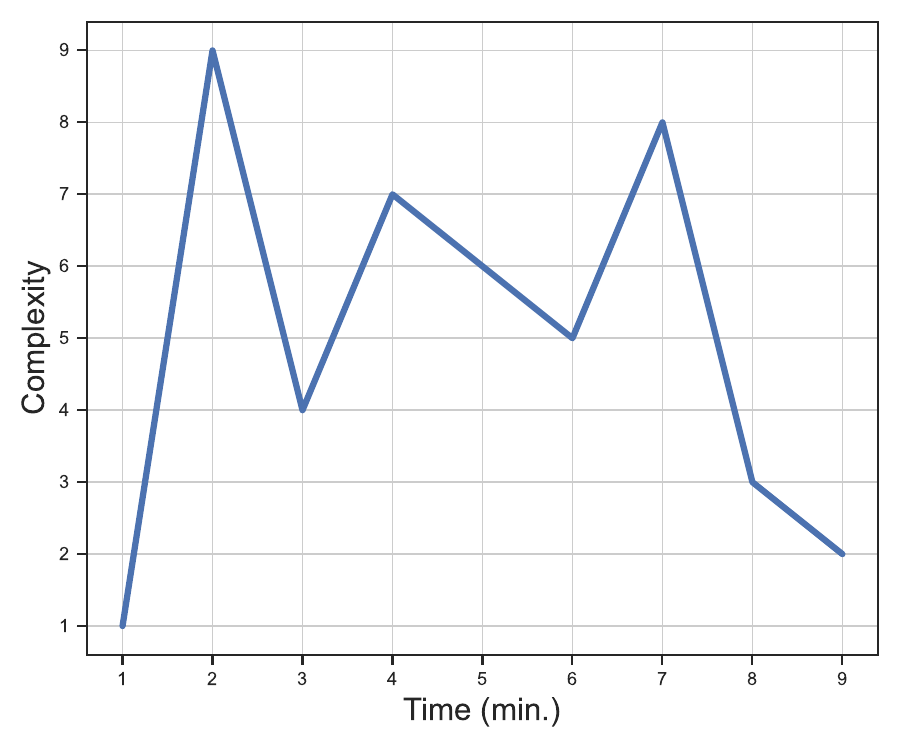}
    \label{fig:seq3}}
    \caption{Sample of complexity levels in 9-minute sessions.}
    \label{fig:complexityTrend}
\end{figure}

We created 9 one-minute complexity levels by modifying these parameters. These mini-experiments with varying complexity levels were then arranged serially to make 9-minute sessions. One session was created in linear `ascending complexity' sequence, while 8 other sessions were created by randomly concatenating the 9 levels of complexity using a Latin square approach. Each participant completed the ascending complexity experiment, followed by 3 of the 8 randomly arranged experiments. A sample of four 9-minute sessions with varying complexities is shown in Figure \ref{fig:complexityTrend}, where the first session (top-left) is the linear ascending complexity session.

\subsection{Ground-truth Cognitive Load Scores}
\begin{figure*}
	\centering
	\includegraphics[width=\linewidth]{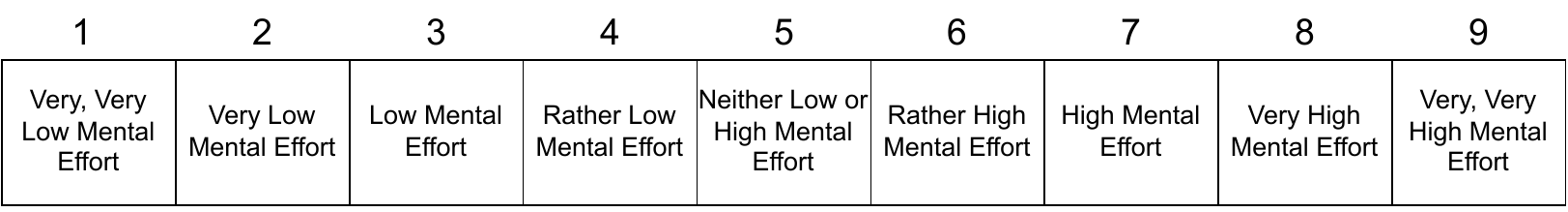}
	\caption[Cognitive load scale.]{Cognitive load levels on the 9-point Likert rating scale.}
	\label{fig:scale}
\end{figure*}

For obtaining the ground-truth scores which can be used as outputs for machine learning models, we used a commonly used 9-point Likert rating scale for cognitive load ratings as introduced in \cite{paas1992training}. This scale ranges from very, very low mental effort (represented by 1) to very, very high mental effort (represented by 9). Participants were told to verbalize their cognitive load according to this scale at 10-second intervals. The 9-point scale is shown in Figure \ref{fig:scale}.

\subsection{Sensors and Placement}
We recorded four modalities namely ECG, EDA, EEG, and gaze data when participants were performing the MATB-II experiments. For ECG, we used a commercially available wearable sensors, Shimmer\footnote{https://shimmersensing.com/product/consensys-bundle-development-kit/ [Accessed:
2022-05-21]}, with which we recorded 3-channel ECG. Sensors comprise five electrodes, of which four electrodes were attached to the clavicle and lower abdomen, and the fifth electrode was connected to the right of the sternum as a reference electrode (see Figure \ref{fig:sensorsplacement}). From these five electrodes the following three signals were collected\footnote{https://shimmersensing.com/wp-content/docs/support/documentation/
ECG\_User\_Guide\_Rev1.12.pdf [Accessed:2022-05-21]} at a sampling frequency of 512 Hz:

\begin{enumerate}
    \item{LA-RA: the ECG vector signal measured from the right arm (RA) position to the left arm (LA) position,}
    \item{LL-RA: the ECG vector signal measured from the RA position to the left leg (LL) position,}
    \item{Vx-RA: the ECG vector signal measured from the Wilson's Central Terminal (WCT) voltage to the Vx position (in our case, sternum). }
\end{enumerate}

The WCT voltage is defined as an artificially constructed virtual reference potential for ECG. It is obtained by taking the average of RA, LA, and LL, and it is assumed that during the cardiac cycle it is steady with negligible amplitude  \cite{moeinzadeh2018modern}. These three signals were recorded and transferred by the electronic module to a laptop for storage. The electronic module was worn on the lower chest using a strap and cradle.

\begin{figure}[t!]
 	\centering
 	\begin{subfigure}{0.4\linewidth}
 		\centering
 		\includegraphics[width=0.8\linewidth]{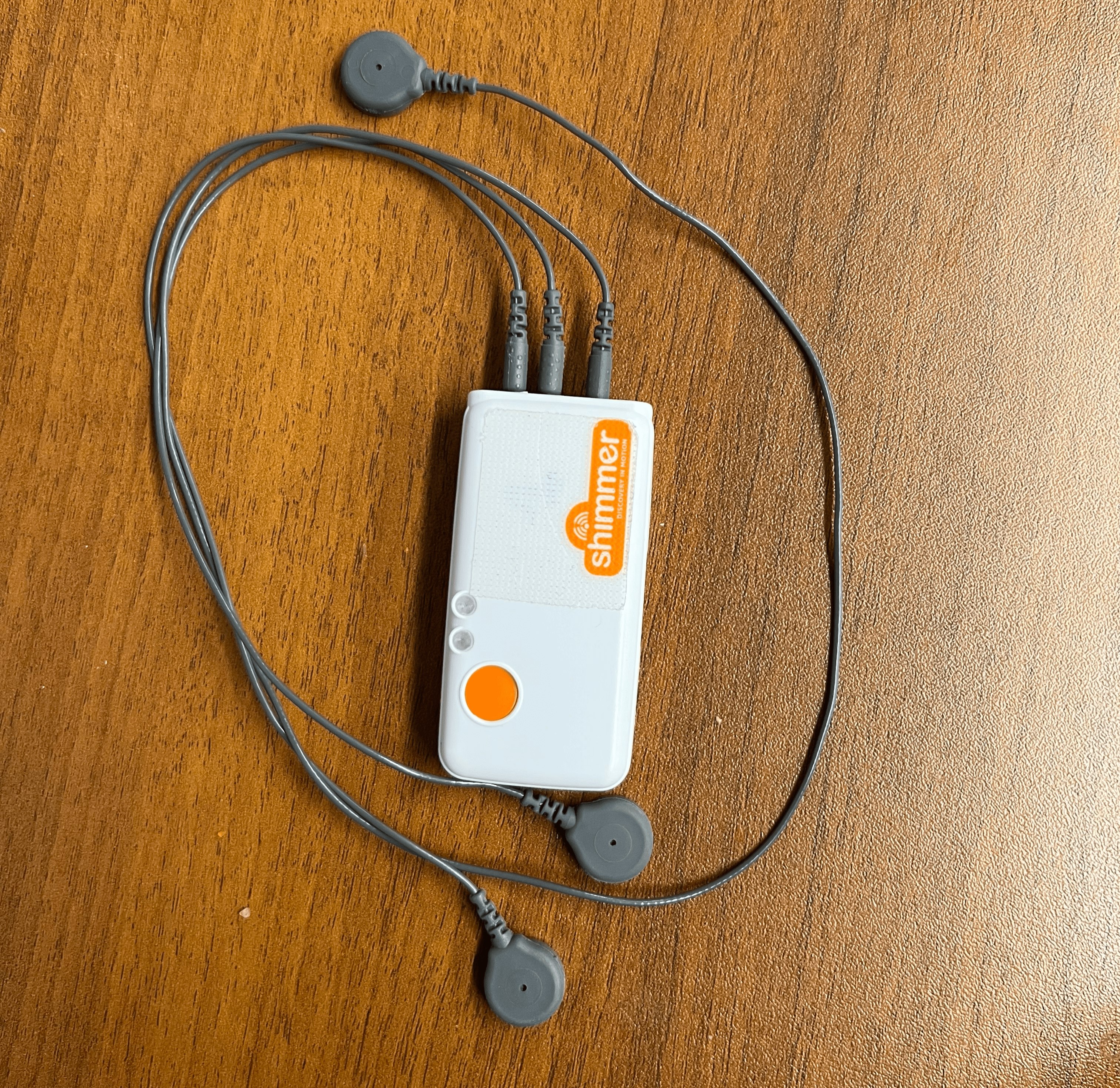}
 		\caption{}
 		\label{fig:shimmer_ecg}
 	\end{subfigure}
 	\begin{subfigure}{0.4\linewidth}
 		\centering
 		\includegraphics[width=0.8\linewidth]{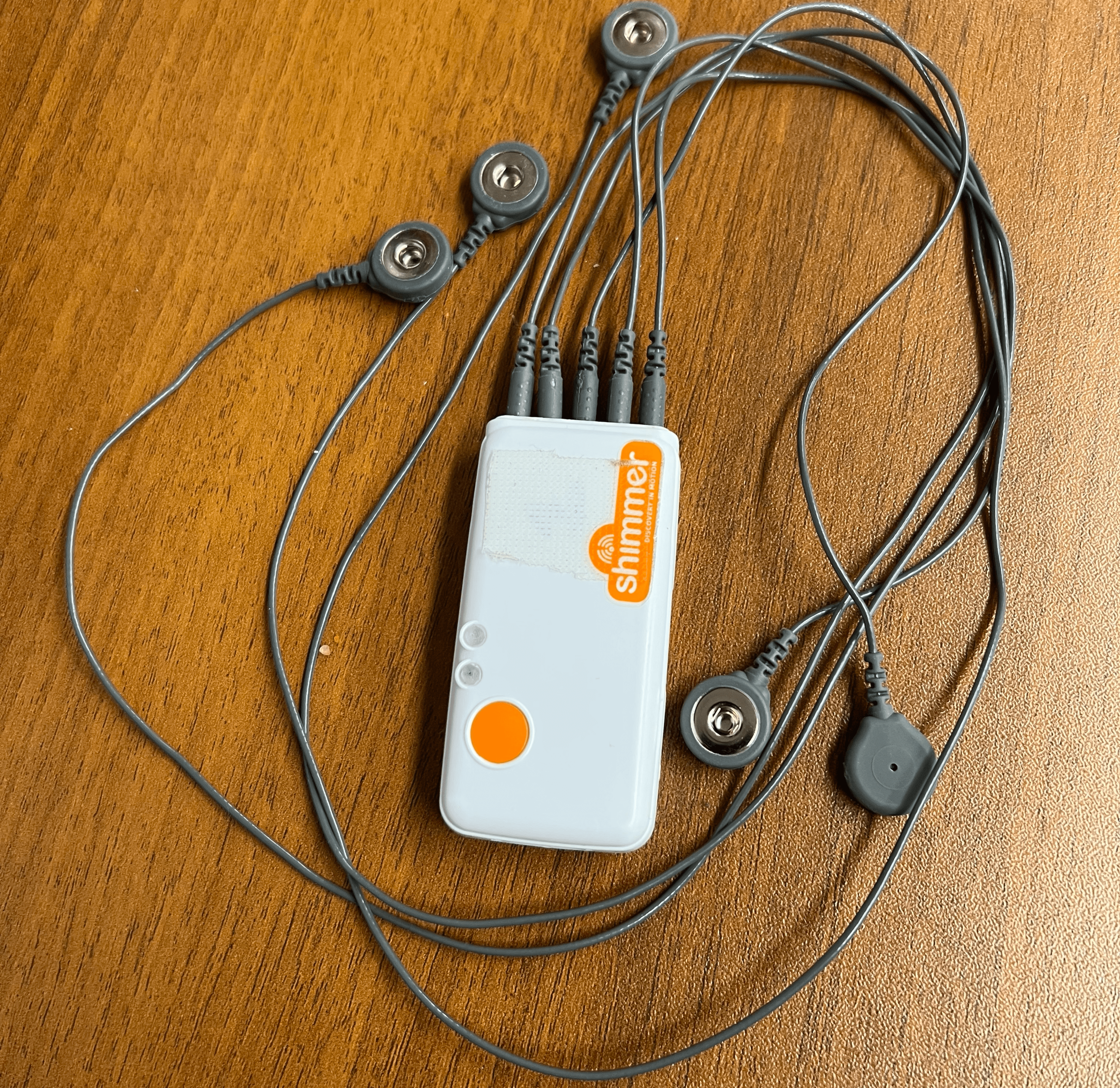}
 		\caption{}
 		\label{fig:shimmer_eda}
 	\end{subfigure}
 	
 	\begin{subfigure}{0.4\linewidth}
 		\centering
 		\includegraphics[width=0.8\linewidth]{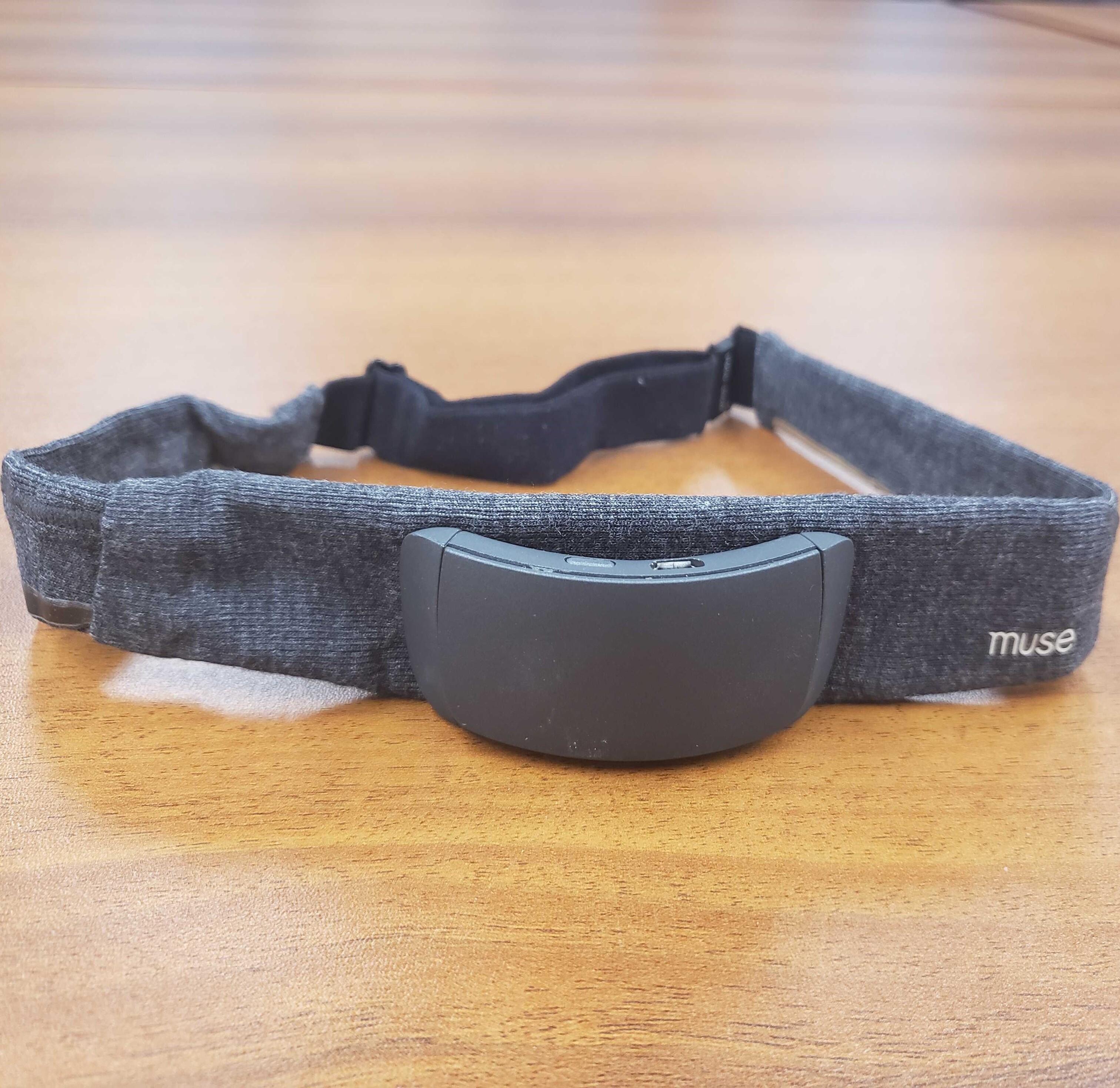}
 		\caption{}
 		\label{fig:museband}
 	\end{subfigure}
 	\begin{subfigure}{0.4\linewidth}
 		\centering
 		\includegraphics[width=0.8\linewidth]{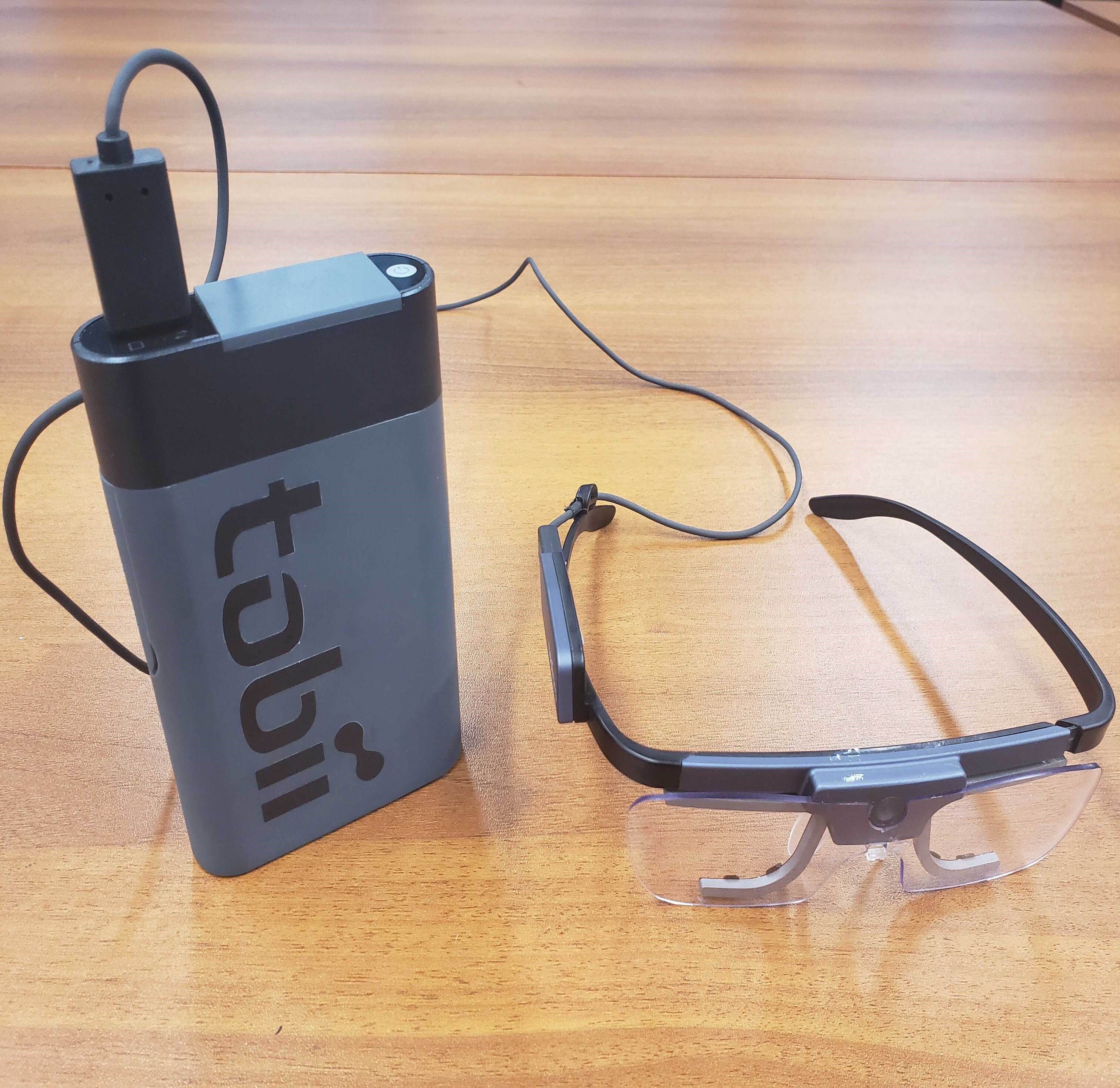}
 		\caption{}
 		\label{fig:tobiiproglasses}
 	\end{subfigure}
 \caption{(a) Shimmer ECG, (b) Shimmer EDA, (c) Muse-S band, and (d) Tobii-Pro 2 glasses}
 \label{fig:sensors}
 \end{figure}

For EDA, we used the Shimmer EDA wearable device\footnote{https://shimmersensing.com/product/consensys-bundle-development-kit/ [Accessed:
2022-05-21]}. As shown in Figure \ref{fig:sensorsplacement}, two sensor electrodes were attached to the index and the middle fingers of the dominant hand using electrode straps. The EDA signal was captured at a sampling frequency of 128 Hz. The electronic module was worn on the wrist of the subject using a strap. The Shimmer ECG and EDA devices are shown in Figures \ref{fig:shimmer_ecg} and \ref{fig:shimmer_eda}.

For collecting EEG, the Muse S Headband\footnote{https://choosemuse.com/muse-s/ [Accessed: 2022-05-21]}, depicted in Figure \ref{fig:museband}, was used. The device was worn on the forehead, and recorded data from four channels: AF7, AF8, TP9 and TP10\footnote{https://choosemuse.com/what-it-measures/ [Accessed: 2022-06-04]}, as shown in Figure \ref{fig:sensorsplacement}. Channels AF7 and AF8 collected data from the forehead, while TP9 and TP10 collected data from the back of the ears. We used an additional conductive gel to maintain better contact between the skin and the electrodes. The raw EEG signal was captured at a sampling frequency of 256 Hz. 

Finally, Tobii Pro Glasses 2\footnote{https://www.tobiipro.com/product-listing/tobii-pro-glasses-2/ [Accessed: 2022-05-21]}, shown in Figure \ref{fig:tobiiproglasses}, were used to record the direction and velocity of the gaze. The glasses consist of two units, a head unit and a recording unit. The gaze-tracking data was recorded at a sampling rate of 50 Hz.
\textcolor{black}{All the data has been collected and synchronized using the iMotion\footnote{https://imotions.com/} software}

\subsection{Experiment Protocol}
Ethics approval was obtained from the Queen's University Health Sciences and Affiliated Teaching Hospitals Research Ethics Board (HSREB), application number: 6025505. Participants gave informed written consent and received instructions about the experiment and the procedure 48 hours before the day of the experiment. Each participant was asked to complete a pre-participation questionnaire within 24 to 12 hours of the participation session that included several items such as quality of sleep, substance use, etc. After the arrival of each participant, members of the research team briefly explained the goal of the experiment, and introduced the participant to the different sensors. Due to the ongoing COVID-19 pandemic at the time of data collection, to maintain physical distancing, all participants placed the sensors themselves with the guidance provided by the data collection team. After the sensors were set up, participants were led to the experiment area. The experiment area comprised the following elements:
\begin{enumerate}
	\item{The participant desk and chair;}
	\item{Computer with 27-inch Dell monitor to display MATB-II software;}
	\item{Mouse and Joystick to enable participant’s responses to the MATB-II tasks;}
	\item{Headset (if required) to hear computer audio during the experiment;}
	\item{Printed participant manual comprising cognitive load scale (1-9).}
\end{enumerate}

\begin{figure}
	\centering
	\includegraphics[width=.8\linewidth]{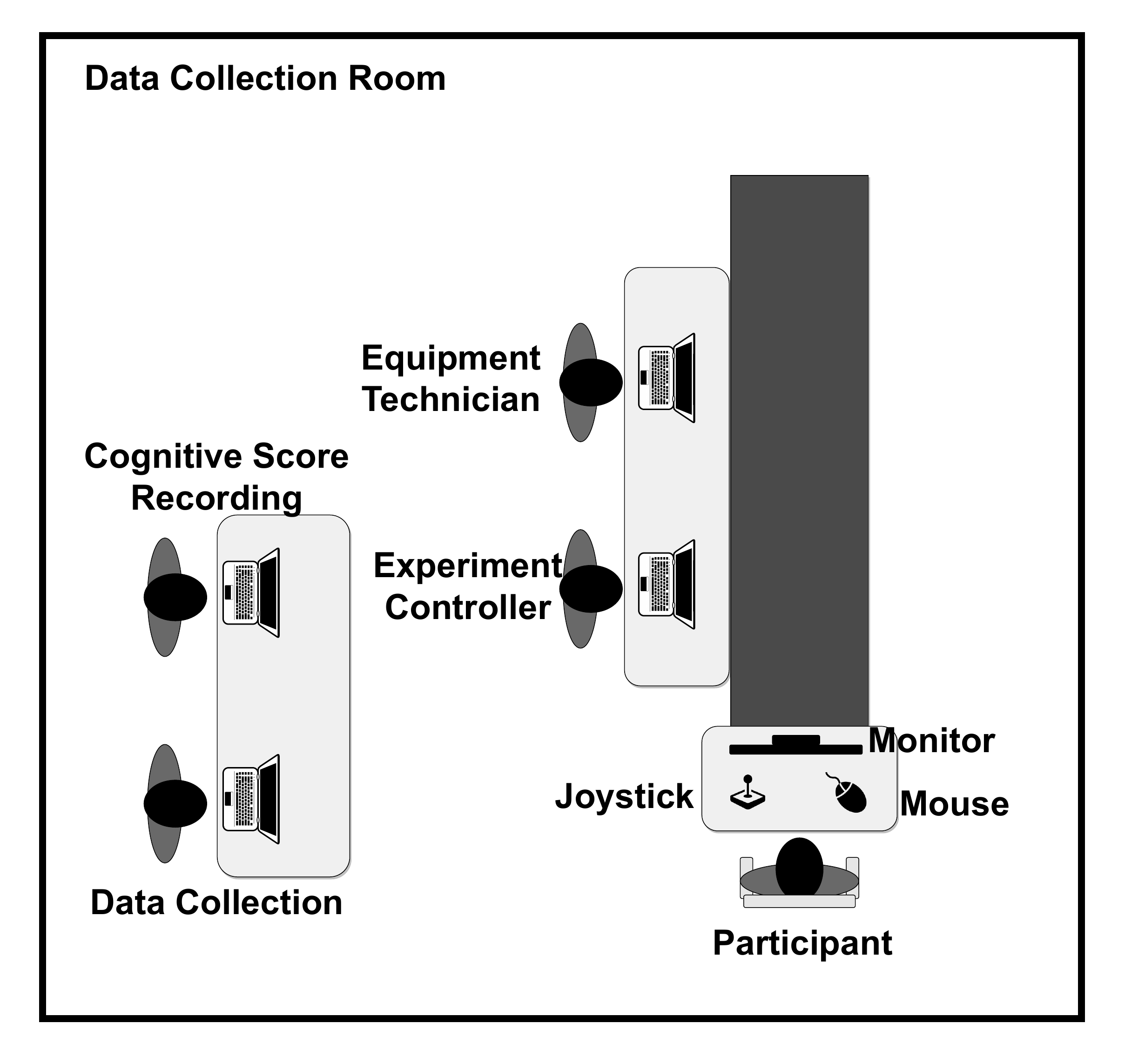}
	\caption{Experiment setup scheme}
	\label{fig:experimentsetting}
\end{figure}

After the participant was seated, the wireless connection to the sensors was established using a remote computer (Dell Alienware 15 laptop) for data capture via blacktooth. An additional blacktooth adapter was used for better connectivity and to reduce data packet loss. Then a member of the research team provided the participant with a guided walkthrough of the experiment, including the purpose of research, sequence of events, expected time commitments, and any procedural tips that could help the participant navigate the process. The layout of the experiment area is shown in Figure \ref{fig:experimentsetting}.

Subsequently, each participant went through a 1-minute practice session per task on the MATB-II, followed by another 1-minute practice session for all the tasks simultaneously. Finally, to make sure that the participants had an understanding of the low, medium, and high complexity levels, they were introduced to low, medium, and high complexity levels for another 30 seconds. After the introductory practice sessions, a 3-minute baseline was collected while the participants rested. Participants were then asked to move their eyes left and right, up and down, and blink ten times to record artifacts generated in the EEG from eye movement. 

As mentioned above, each participant was asked to complete four 9-minute MATB-II sessions, the first of which was the linearly ascending complexity session. Participants provided their subjective cognitive load scores verbally during each session by responding to a beep sound after every 10-seconds. After each 9-minute session, each participant was given 2-minutes of resting time. 
At the end of the fourth session, participants completed a debriefing discussion by answering some qualitative questions about the experiment. All participants received a monetary compensation of 70 CAD after the experiment. 

\section{Method} \label{sec:method}

\subsection{Data Preprocessing}
\textcolor{black}{In this section we discuss various pre-processing steps used to filter and clean the recorded data. The raw ECG signal comprises EMG noise, powerline noise, baseline wander, and T-wave interference. We applied a Butterworth bandpass filter with a passband frequency of 5-15 \textit{Hz} \cite{pan1985real}. User-specific z-score normalization was applied to the filtered ECG signal \cite{behinaein2021transformer, bhatti2021attentive}.} 

\textcolor{black}{We removed high-frequency noise from the raw EDA signal using a lowpass filter with a cut-off frequency of 3 \textit{Hz}. Like ECG, we also applied user-specific z-score normalization to the filtered EDA signal. Further, a highpass filter with a cut-off frequency of 0.05 \textit{Hz} decomposed the filtered EDA signal into skin conductance level, also known as tonic level, and skin conductance response, also known as the phasic response \cite{schmidt2018introducing, bhatti2021attentive}.} 

\textcolor{black}{As mentioned in the previous section, the raw EEG signal has four channels, namely TP9, AF7, AF8, and TP10. These four channels are filtered using a Butterworth lowpass filter with a passband frequency of 0.4-128 Hz. Since powerline noise frequently contaminates the EEG signal, we utilized a notch filter at 60 Hz with a quality factor of 30 to remove it from each channel. Then, we performed user-specific z-score normalization to each filtered channel.}

{\renewcommand{\arraystretch}{1.7}
\begin{table}[!t]
\caption{Extracted handcrafted affective features from each Modality for classical machine learning algorithms.}
\label{tbl:extracted-features}
\centering
\begin{tabular}{cp{0.1\linewidth}p{0.6\linewidth}}
\midrule[.005\linewidth]
\textbf{Modality} & \multicolumn{2}{c}{\textbf{Extracted features}}\\
\midrule[.005\linewidth]
\multirow{11}{*}{ECG} & \multirow{6}{0pt}{Time Domain} & Min HR, Min HRV, Min ECG, Max HR, Max HRV, Max ECG, Mean HR, Mean HRV, Mean ECG, SD HR, SD HRV, SD, ECG, RMS HRV, \% HRV $>$ 50 ms, \% HRV $>$ 20 ms, IQR HRV, IQR ECG, MAD HRV, MAD ECG, Skewness ECG, kurtosis ECG, entropy ECG, AUC$^2$ ECG \\ \cmidrule{2-3}

& \multirow{5}{0pt}{Freq. Domain} & Peak frequency - ULF, VLF, LF, HF\\
& & Absolute power - ULF, VLF, LF, HF\\
& & Normalized power of LF \& HF\\
& & LF/HF ratio, Total power \\
\midrule
\multirow{4}{*}{EDA} & \multirow{4}{0pt}{Time Domain} & Min., Max., Mean, SD - EDA, Phasic, Tonic, Amplitude Phasic, Height Phasic, Recovery Time, Rise Time\\
& & \# of Peaks in Phasic \\
& & IQR, MAD - EDA, Phasic, Tonic\\  
& & Skewness, Kurtosis, Entropy, AUC$^2$ - EDA, Phasic, Tonic\\  
\midrule
\multirow{6}{*}{EEG} & \multirow{5}{0pt}{Time Domain} & Min., Max., Mean, Median - EEG, FFT\\
& & Spectral Entropy\\
& & Hjorth mobility and complexity\\
& & Lempel-Ziv Complexity\\
& & Higuchi fractal dimension\\ \cmidrule{2-3}
& \multirow{2}{0pt}{Freq. Domain} & Min., Max., Mean, Median, Abs. Power of each Frequency Band\\
\midrule
\multirow{2}{*}{Gaze} & \multirow{5}{0pt}{Time Domain} & Min., Max., Mean - Pupil Diameter (Left, Right), Blink, Fixation, Fixation Dispersion, Saccade, Saccade Amplitude, Saccade Peak Velocity, Saccade Peak Acceleration, Saccade Peak Deceleration, Saccade Direction\\
& & Number of Blinks, Fixations, \& Saccades\\
\bottomrule[.005\linewidth]
\end{tabular}
\end{table}
}

{\renewcommand{\arraystretch}{1.4}
\begin{table} 
\begin{center}
\caption[Architectural details]{The architectural details (filter size, no. of filters, and stride) of the CNN network.}
\label{tbl:archdetails}
\scriptsize
\begin{tabular}{cc} 
\midrule[.005\linewidth]
\multirow{1}{*}{\textbf{Block No.}} & \textbf{Backbone Architecture with VGG Blocks} \\ \midrule[.005\linewidth]

\multirow{2}{*}{Block 1} & \multirow{2}{*}{$\begin{array}{cc}
    \text{Conv1D, 64, 32, 1}\\
    \text{Conv1D, 64, 32, 3}\\
\end{array}$} \\
& \\ \midrule
\multirow{2}{*}{Block 2} & \multirow{2}{*}{$\begin{array}{cc}
    \text{Conv1D, 32, 64, 1}  \\
    \text{Conv1D, 32, 64, 3}\\
\end{array}$} \\
& \\ \midrule

\multirow{2}{*}{Block 3} & \multirow{2}{*}{$\begin{array}{cc}
    \text{Conv1D, 17, 128, 1}  \\
    \text{Conv1D, 17, 128, 3}\\
\end{array}$} \\
& \\ \midrule

\multirow{2}{*}{Block 4} & \multirow{2}{*}{$\begin{array}{cc}
    \text{Conv1D, 7, 256, 1}  \\
    \text{Conv1D, 7, 256, 3}\\
\end{array}$} \\
& \\ \midrule

\multicolumn{2}{c}{Fully Connected, 512} \\ \midrule
\multicolumn{2}{c}{Fully Connected, 256} \\ \midrule

\multicolumn{2}{c}{SoftMax, 2} \\ \bottomrule[.005\linewidth]
\end{tabular}
\end{center}
\end{table}}

\begin{figure*}
    \centering
     \begin{subfigure}{0.15\textwidth}
         \includegraphics[width=1\textwidth]{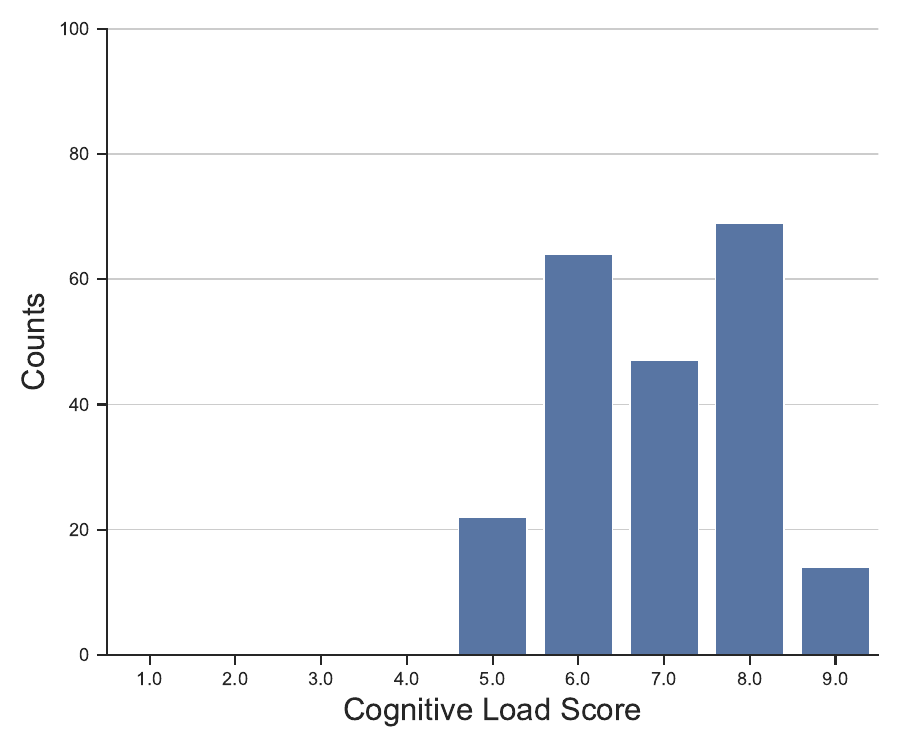}
         \caption{\footnotesize Subject 1}
         \label{fig:sub1}
     \end{subfigure}
     \begin{subfigure}{0.15\textwidth}
         \includegraphics[width=1\textwidth]{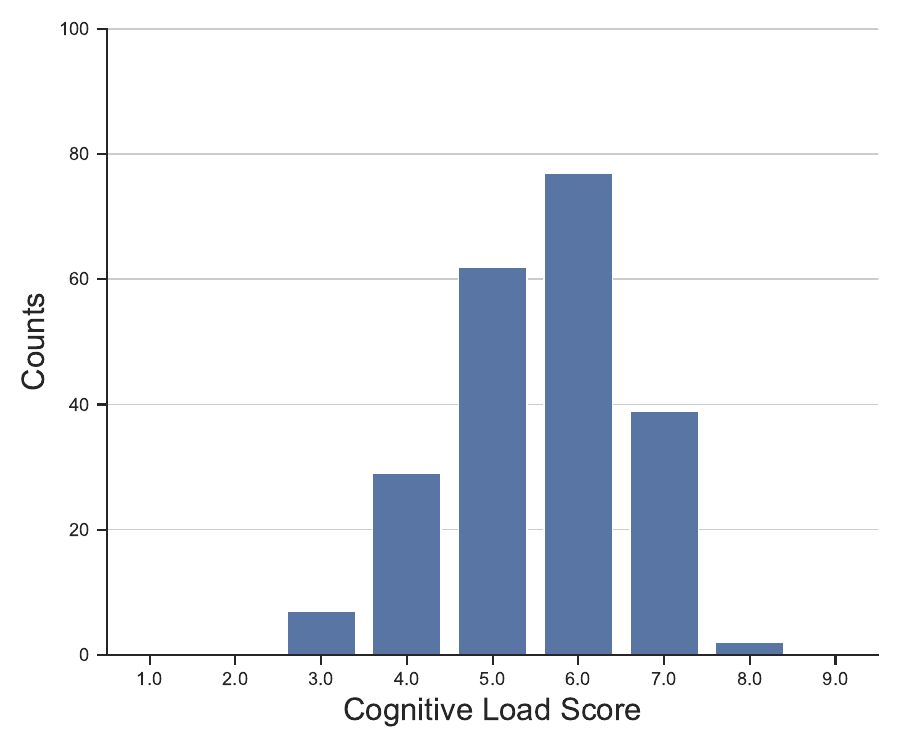}
         \caption{\footnotesize Subject 2}
         \label{fig:sub2}
     \end{subfigure}
     \begin{subfigure}{0.15\textwidth}
         \includegraphics[width=1\textwidth]{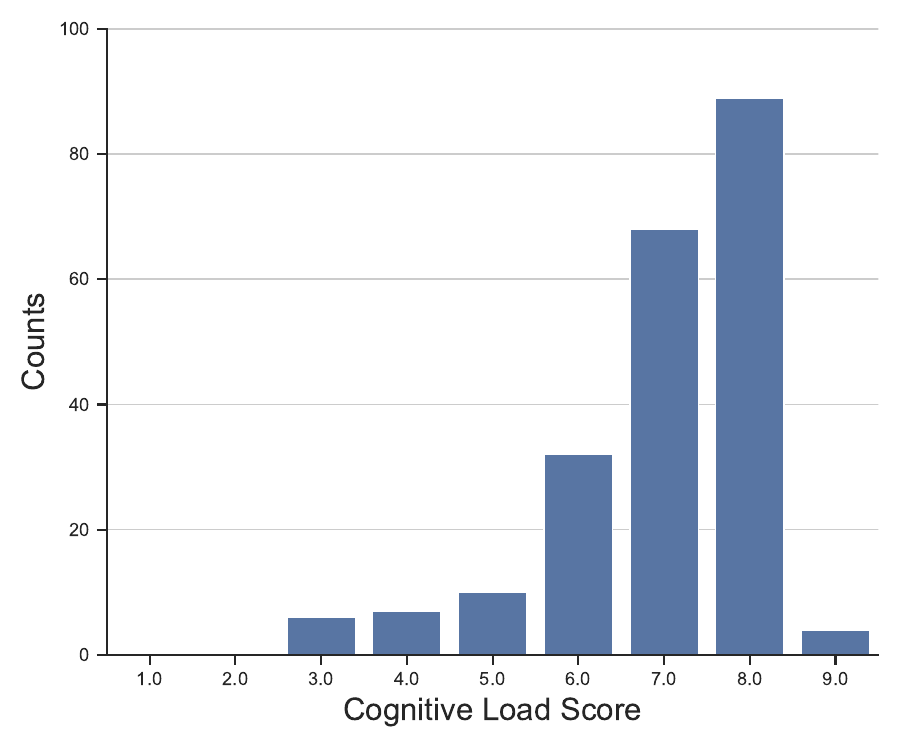}
         \caption{\footnotesize Subject 3}         
         \label{fig:sub3}
     \end{subfigure}
     \begin{subfigure}{0.15\textwidth}
         \includegraphics[width=1\textwidth]{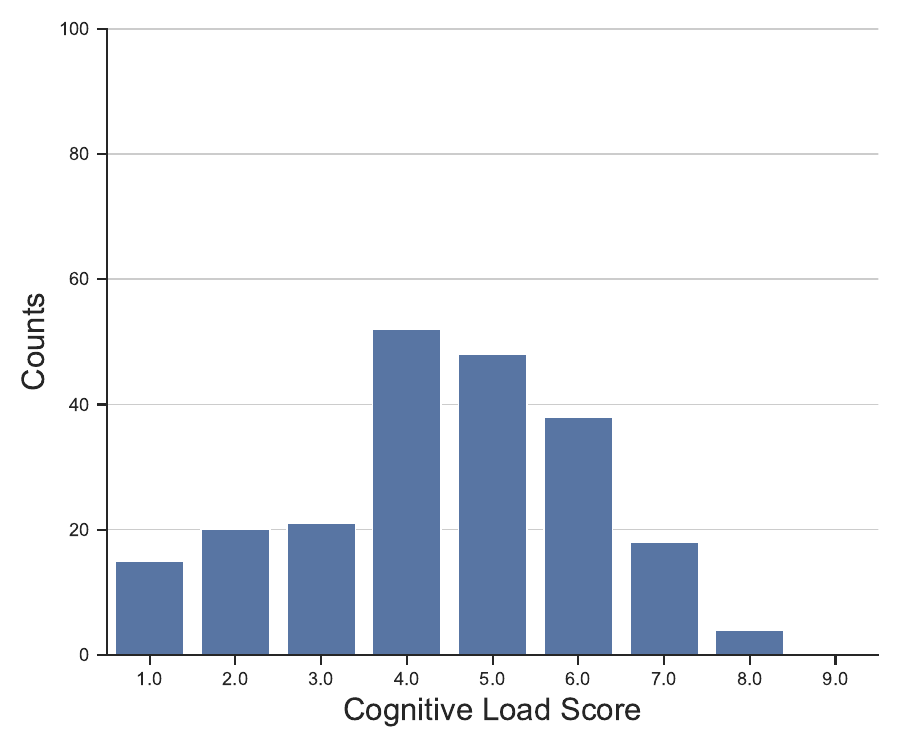}
         \caption{\footnotesize Subject 4}         
         \label{fig:sub4}
     \end{subfigure}
     \begin{subfigure}{0.15\textwidth}
         \includegraphics[width=1\textwidth]{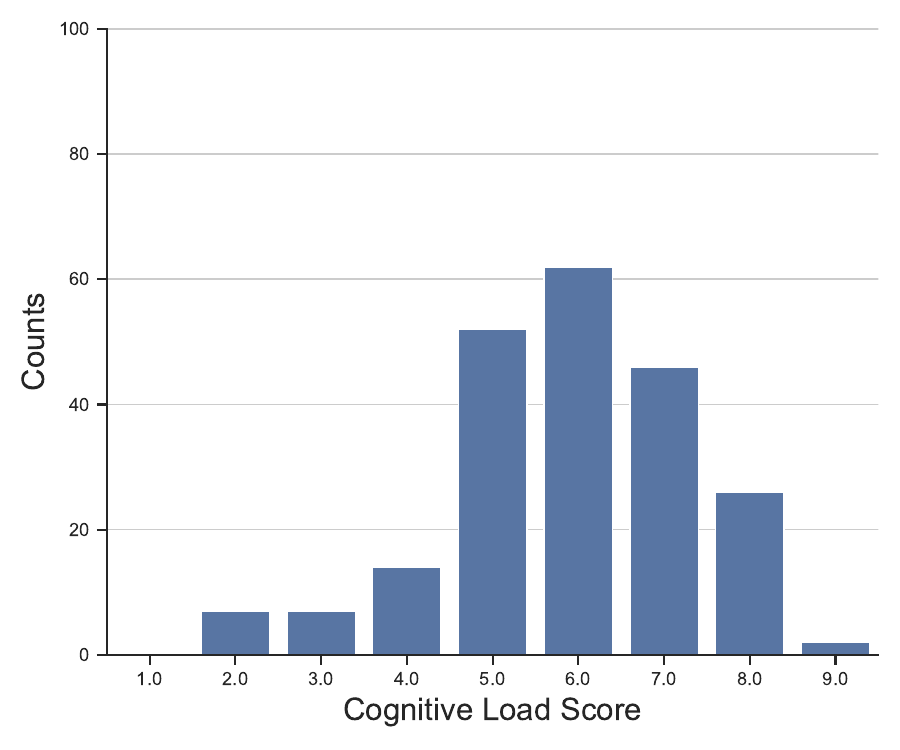}
         \caption{\footnotesize Subject 5}         
         \label{fig:sub5}
     \end{subfigure}
     \begin{subfigure}{0.15\textwidth}
         \includegraphics[width=1\textwidth]{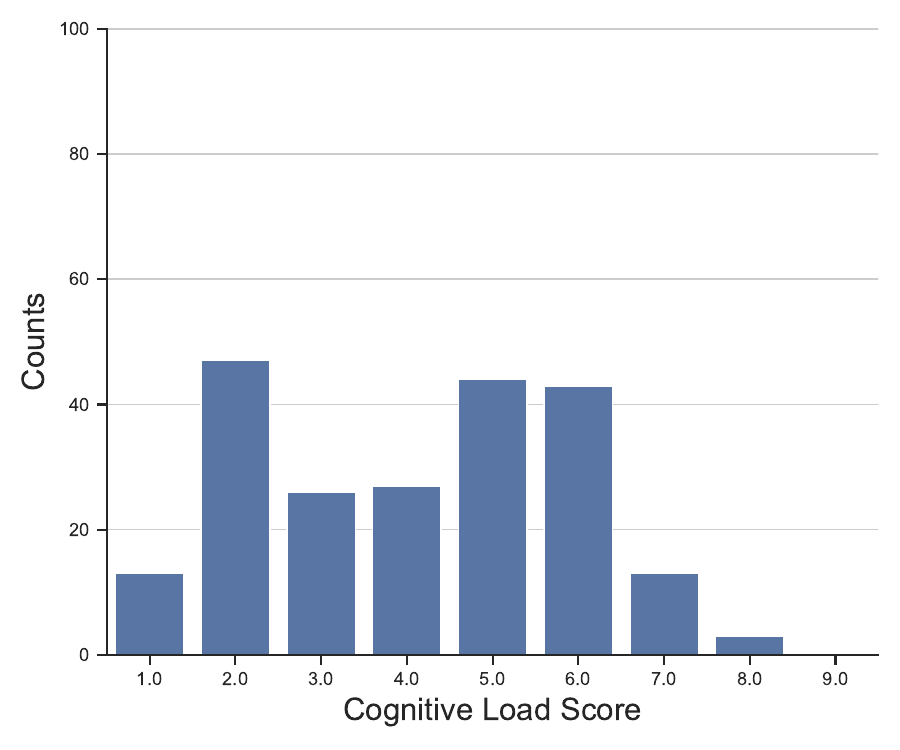}
         \caption{\footnotesize Subject 6}         
         \label{fig:sub6}
     \end{subfigure}
     \begin{subfigure}{0.15\textwidth}
         \includegraphics[width=1\textwidth]{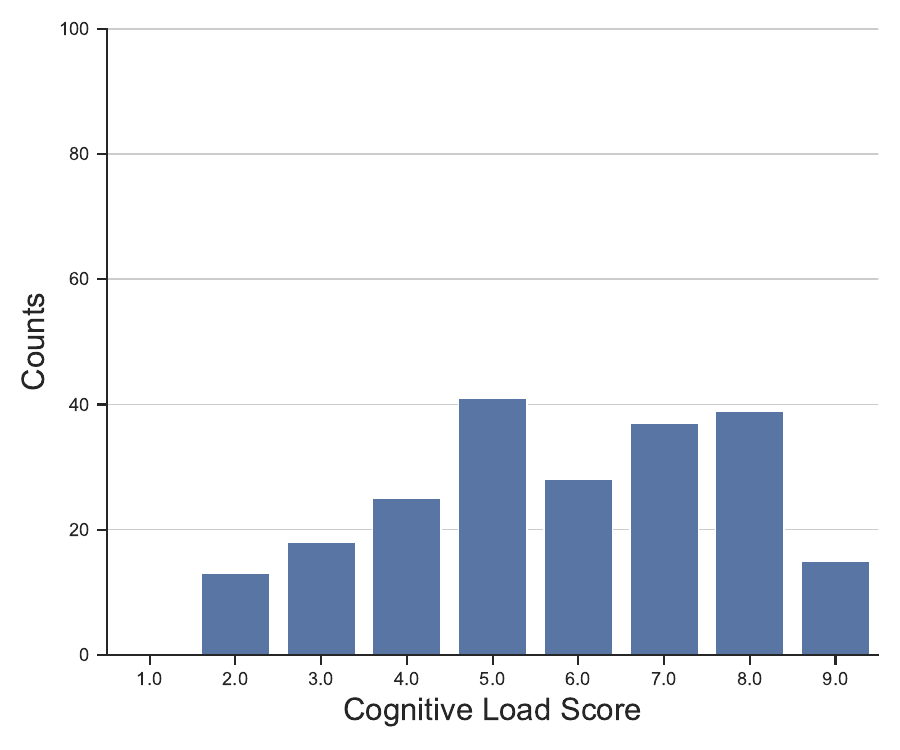}
         \caption{\footnotesize Subject 7}         
         \label{fig:sub7}
     \end{subfigure}
     \begin{subfigure}{0.15\textwidth}
         \includegraphics[width=1\textwidth]{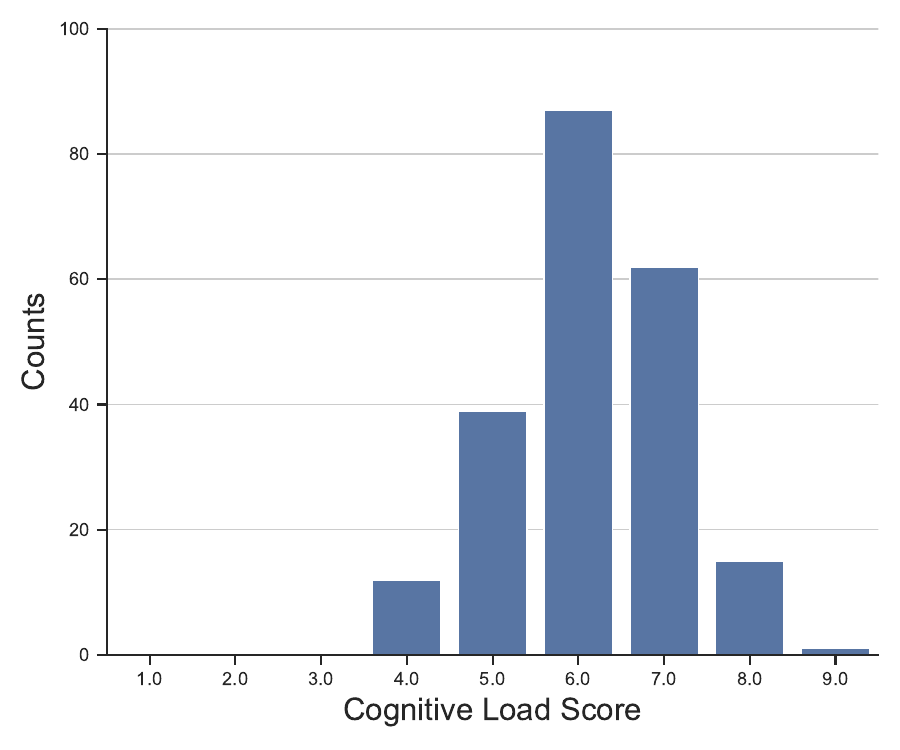}
         \caption{\footnotesize Subject 8}
         \label{fig:sub8}
     \end{subfigure}
     \begin{subfigure}{0.15\textwidth}
         \includegraphics[width=1\textwidth]{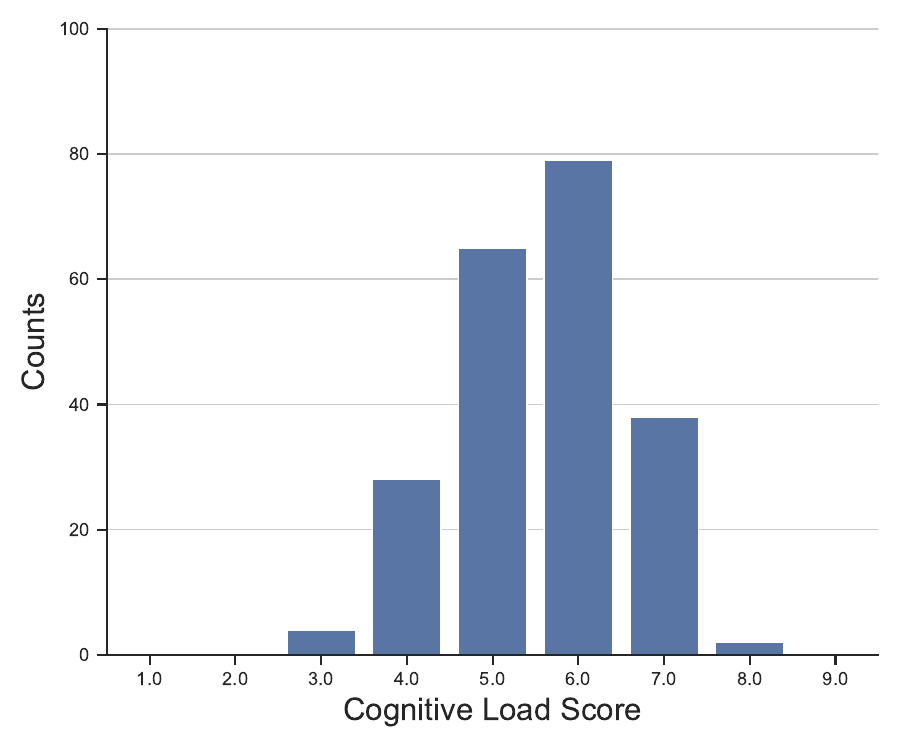}
         \caption{\footnotesize Subject 9}
         \label{fig:sub9}
     \end{subfigure}
     \begin{subfigure}{0.15\textwidth}
         \includegraphics[width=1\textwidth]{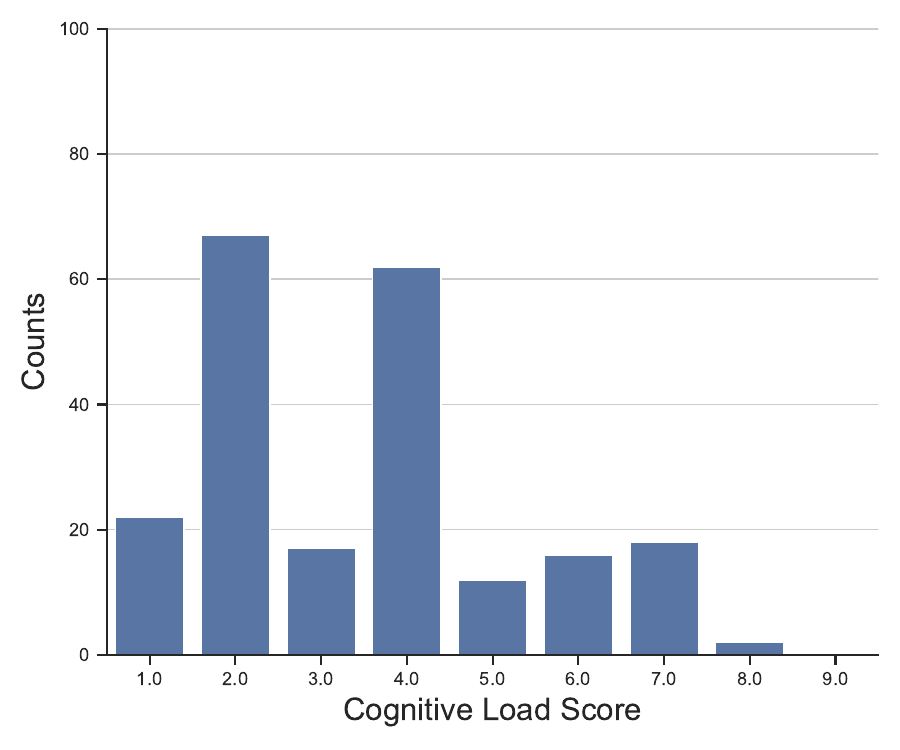}
         \caption{\footnotesize Subject 10}
         \label{fig:sub10}
     \end{subfigure}
     \begin{subfigure}{0.15\textwidth}
         \includegraphics[width=1\textwidth]{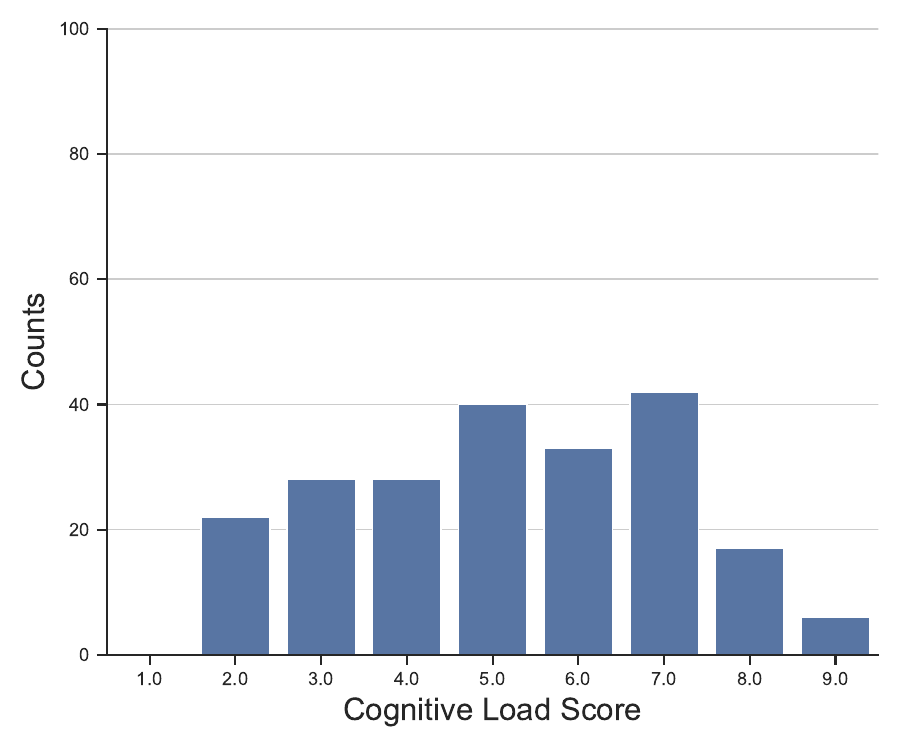}
         \caption{\footnotesize Subject 11}
         \label{fig:sub11}
     \end{subfigure}     
     \begin{subfigure}{0.15\textwidth}
         \includegraphics[width=1\textwidth]{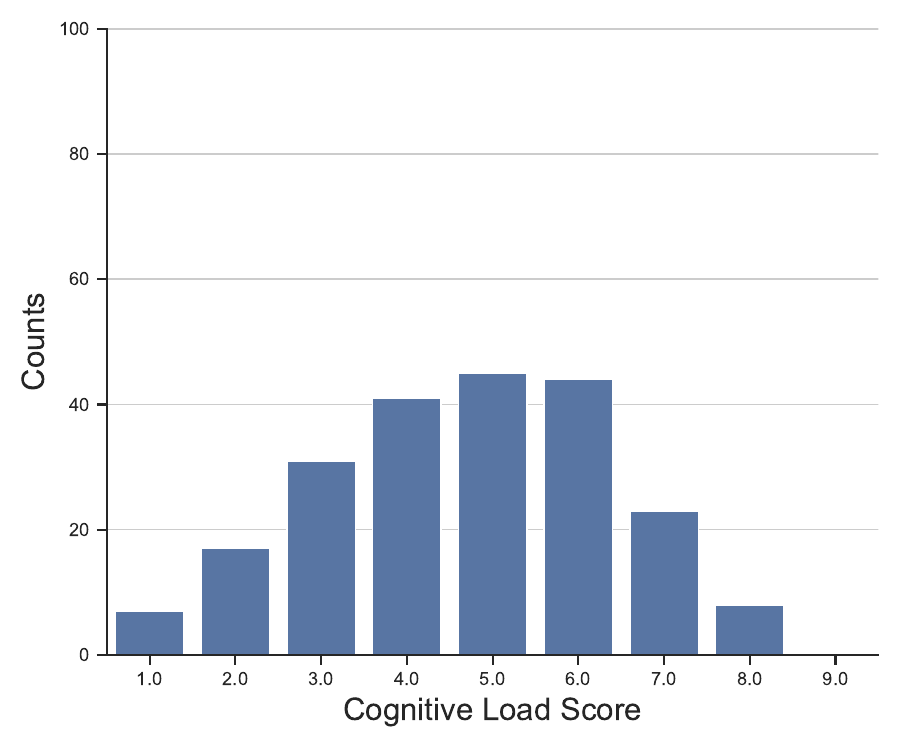}
         \caption{\footnotesize Subject 12}
         \label{fig:sub12}
     \end{subfigure}
     \begin{subfigure}{0.15\textwidth}
         \includegraphics[width=1\textwidth]{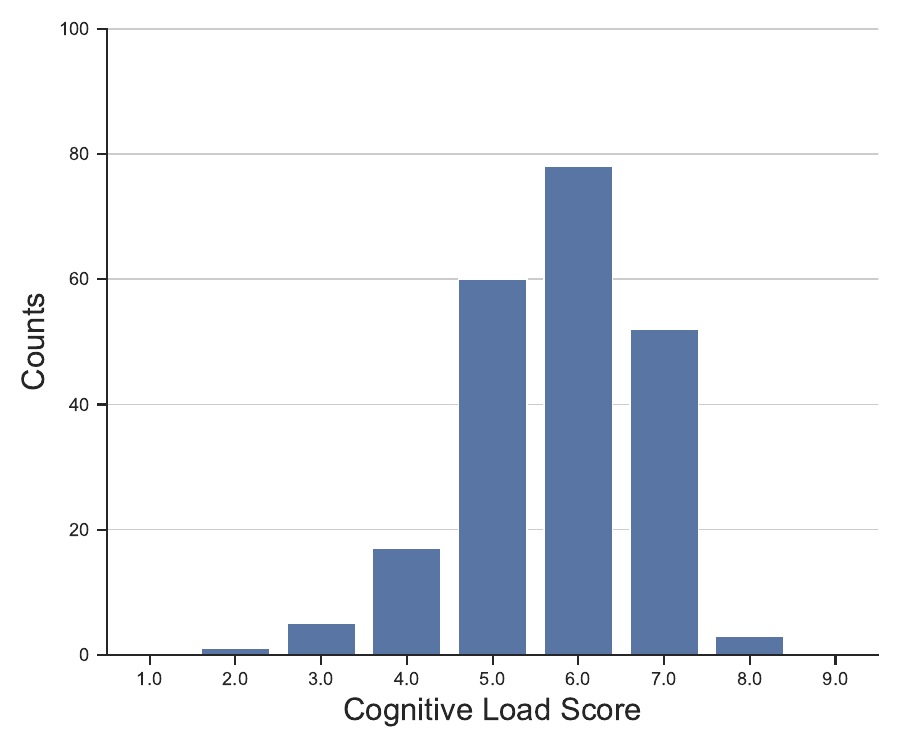}
         \caption{\footnotesize Subject 13}
         \label{fig:sub13}
     \end{subfigure}
     \begin{subfigure}{0.15\textwidth}
         \includegraphics[width=1\textwidth]{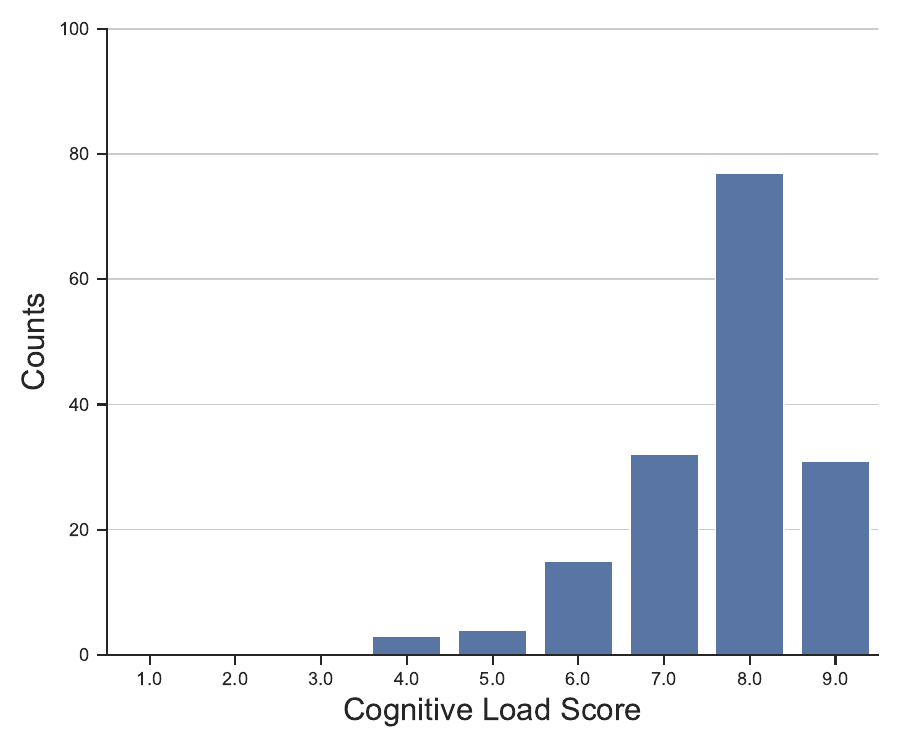}
         \caption{\footnotesize Subject 14}
         \label{fig:sub14}
     \end{subfigure}
     \begin{subfigure}{0.15\textwidth}
         \includegraphics[width=1\textwidth]{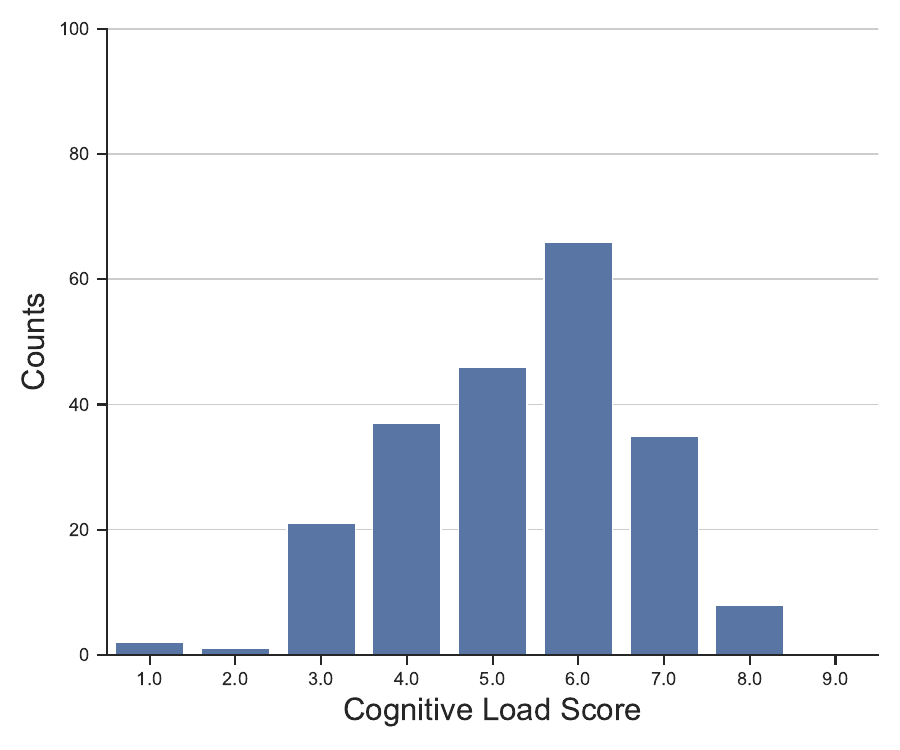}
         \caption{\footnotesize Subject 15}
         \label{fig:sub15}
     \end{subfigure}     
     \begin{subfigure}{0.15\textwidth}
         \includegraphics[width=1\textwidth]{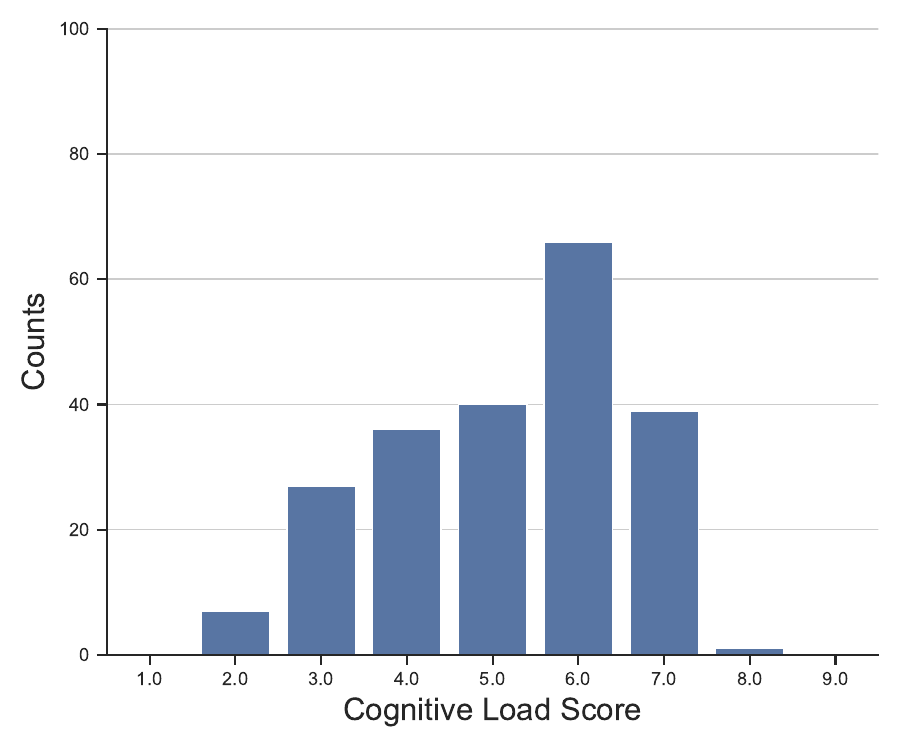}
         \caption{\footnotesize Subject 16}
         \label{fig:sub16}
     \end{subfigure}     
     \begin{subfigure}{0.15\textwidth}
         \includegraphics[width=1\textwidth]{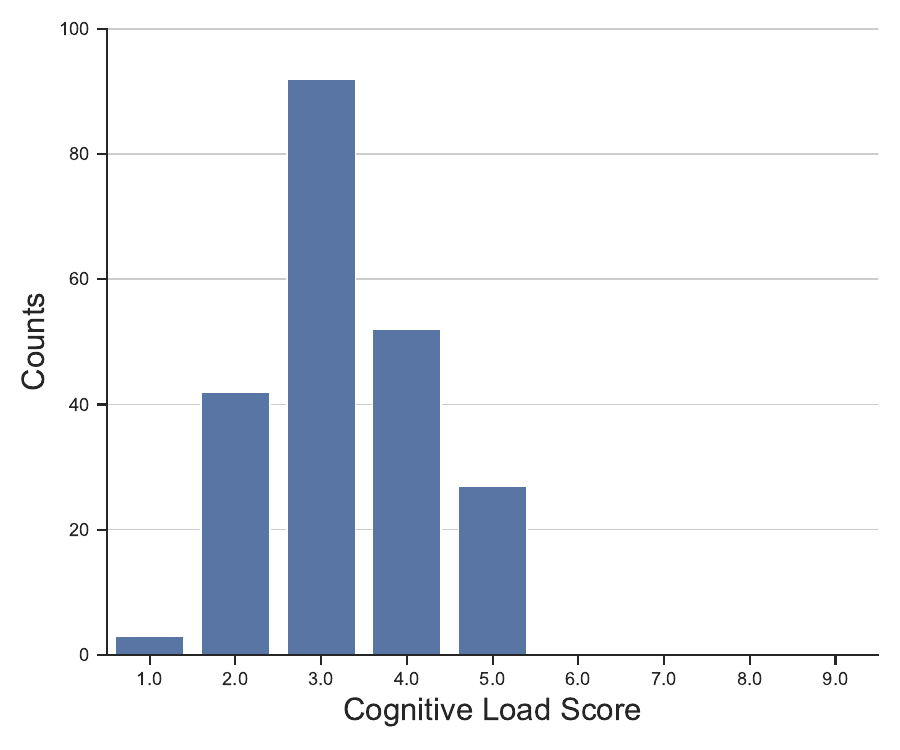}
         \caption{\footnotesize Subject 17}
         \label{fig:sub17}
     \end{subfigure}
     \begin{subfigure}{0.15\textwidth}
         \includegraphics[width=1\textwidth]{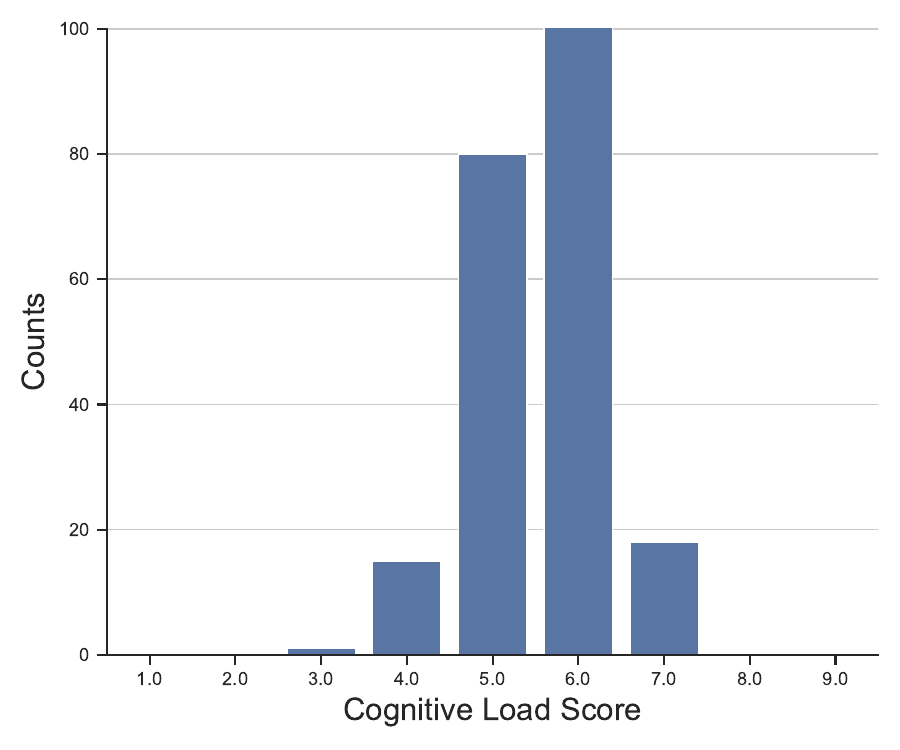}
         \caption{\footnotesize Subject 18}
         \label{fig:sub18}
     \end{subfigure}
     \begin{subfigure}{0.15\textwidth}
         \includegraphics[width=1\textwidth]{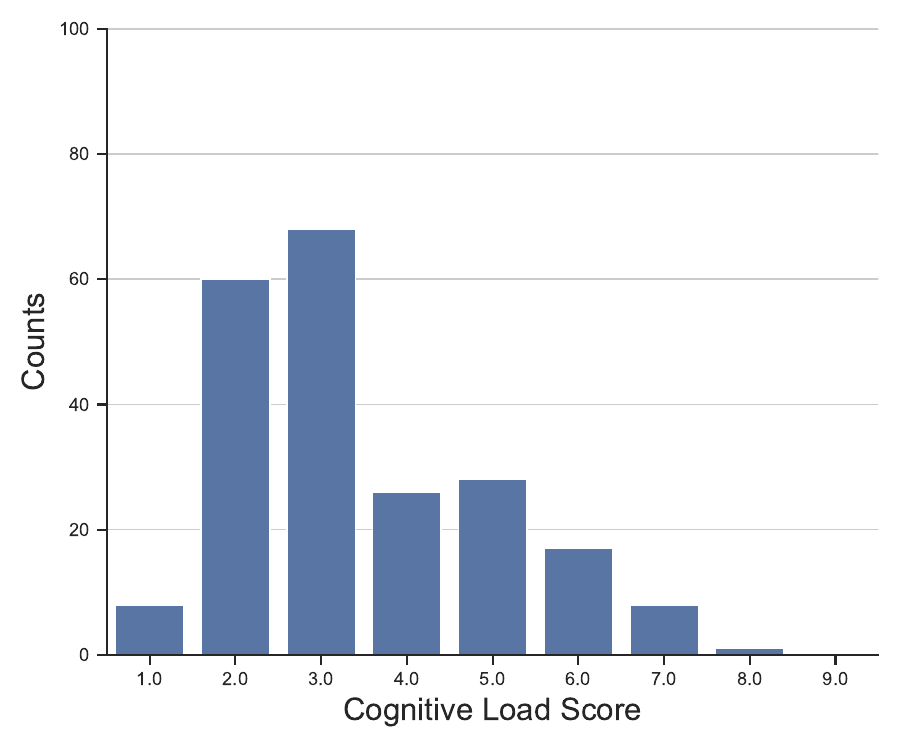}
         \caption{\footnotesize Subject 19}
         \label{fig:sub19}
     \end{subfigure}
     \begin{subfigure}{0.15\textwidth}
         \includegraphics[width=1\textwidth]{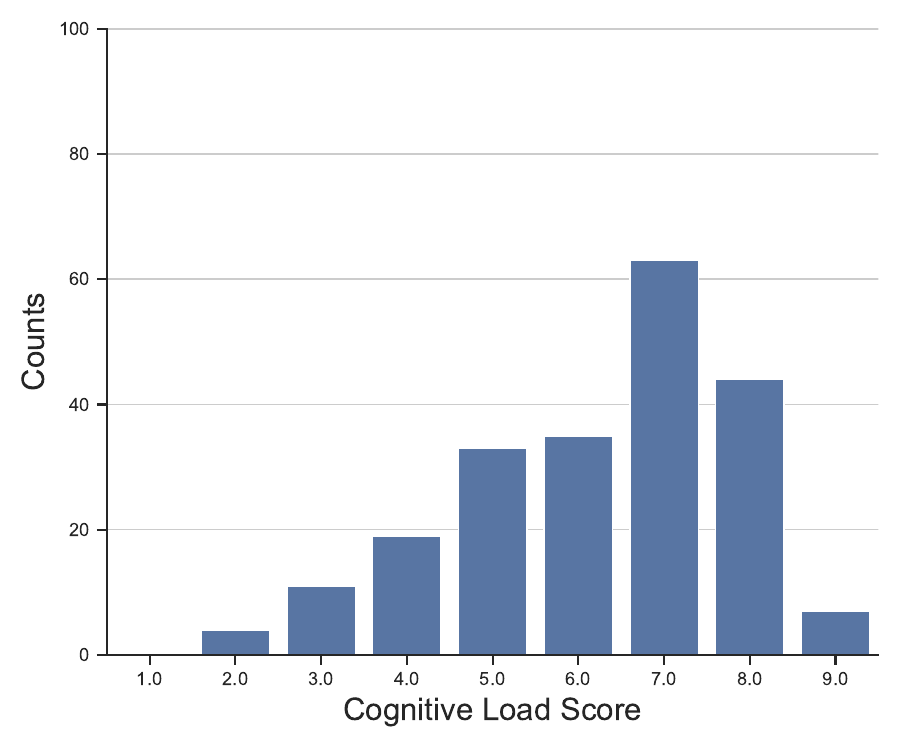}
         \caption{\footnotesize Subject 20}
         \label{fig:sub20}
     \end{subfigure}
     \begin{subfigure}{0.15\textwidth}
         \includegraphics[width=1\textwidth]{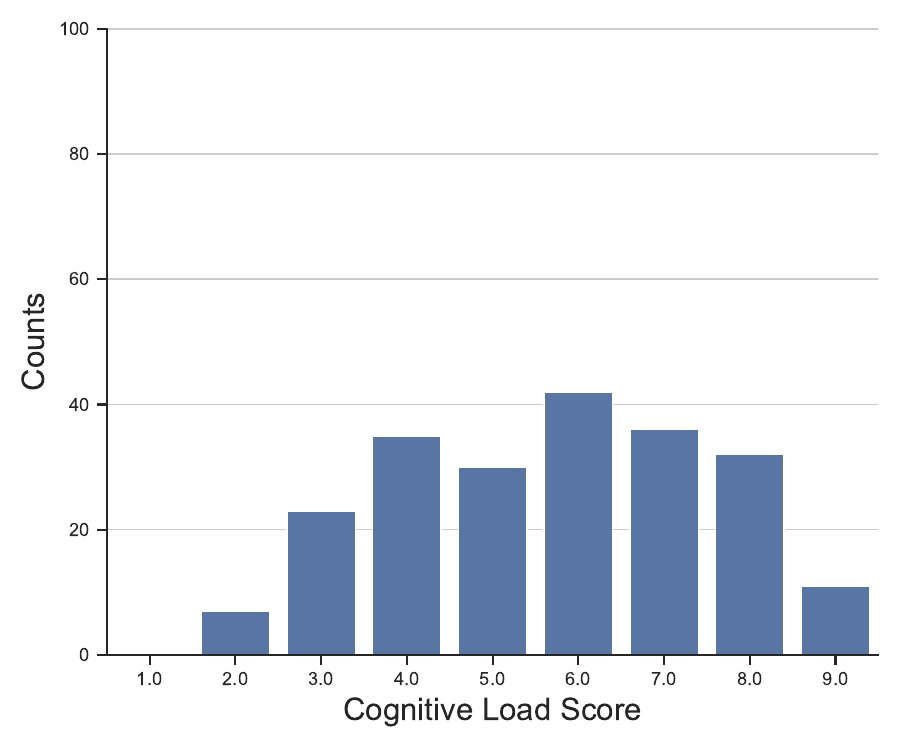}
         \caption{\footnotesize Subject 21}
         \label{fig:sub21}
     \end{subfigure}     
     \begin{subfigure}{0.15\textwidth}
         \includegraphics[width=1\textwidth]{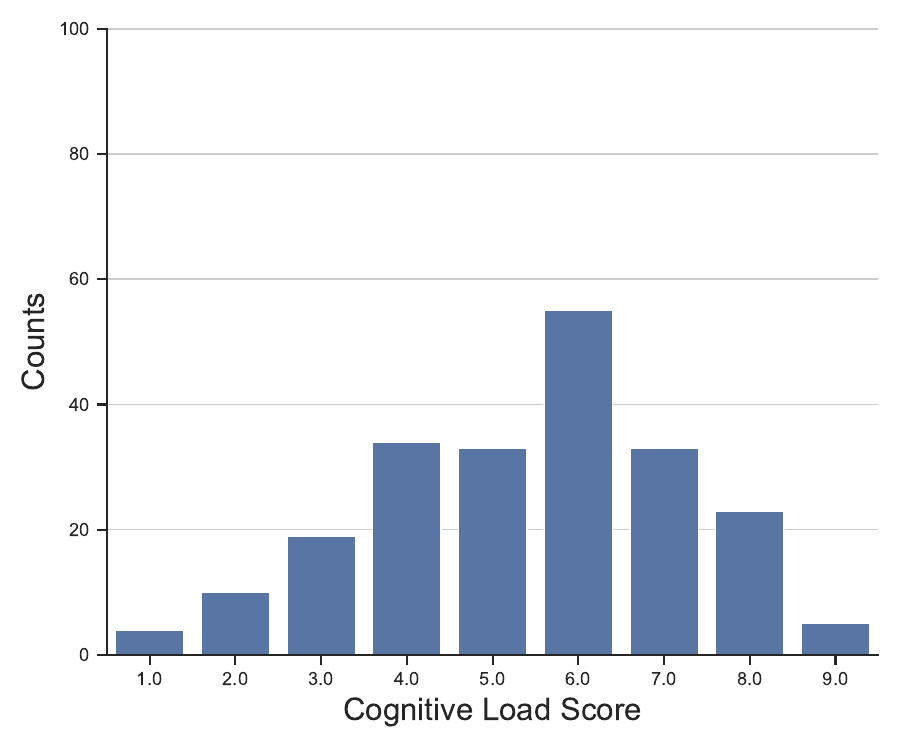}
         \caption{\footnotesize Subject 22}
         \label{fig:sub22}
     \end{subfigure}
     \begin{subfigure}{0.15\textwidth}
         \includegraphics[width=1\textwidth]{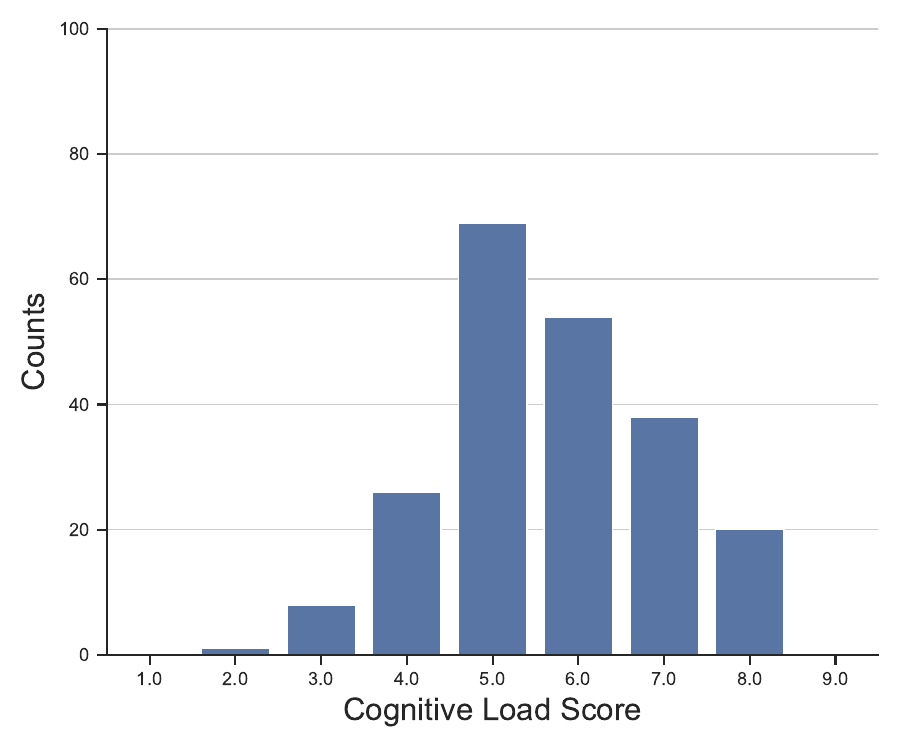}
         \caption{\footnotesize Subject 23}
         \label{fig:sub23}
     \end{subfigure}
     \begin{subfigure}{0.15\textwidth}
         \includegraphics[width=1\textwidth]{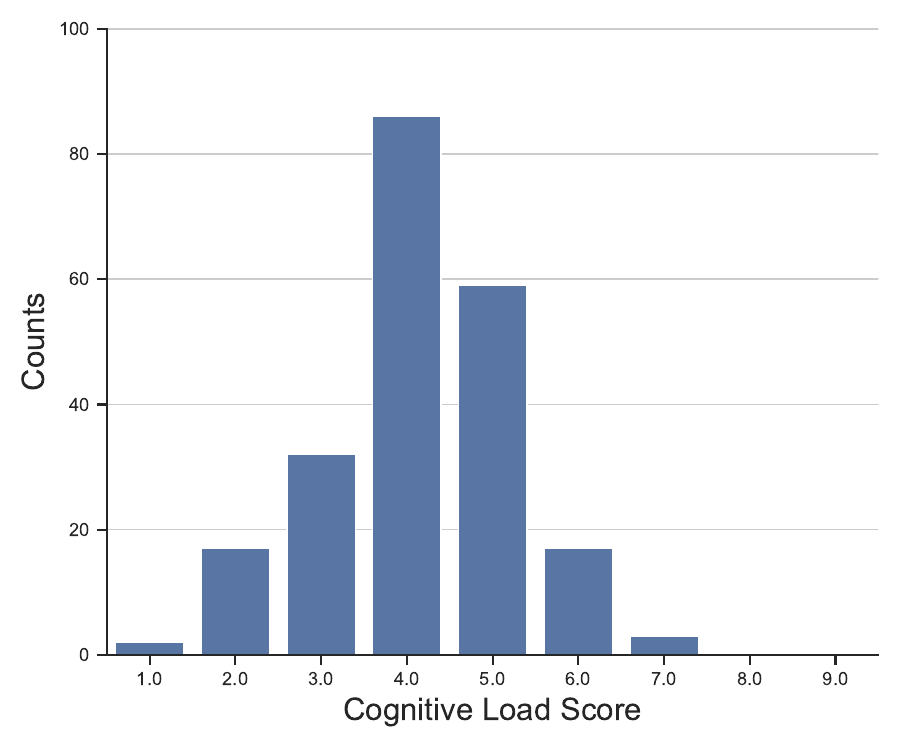}
         \caption{\footnotesize Subject 24}
         \label{fig:sub24}
     \end{subfigure}
     \caption{Distribution of reported cognitive load values for each participant. Self-reported cognitive load levels are shown on the \textit{x} axis, while \textit{y} axis presents the number of times each level has been reported.}
     \label{fig:labeldist}
\end{figure*}

\subsection{Feature Extraction}
\label{subsec:feature_extraction}
After cleaning each signal, the filtered signals are divided into 10-second segments to compute various statistical and modality-specific features from each segment for modelling. A list of all features computed from ECG, EDA, EEG, and Gaze is presented in Table \ref{tbl:extracted-features}, and described below. 

\textbf{ECG Features}: We extracted various statistical, time domain and frequency domain features from 10-second segments of filtered ECG signals. The features included time domain features such as mean, standard deviation, maximum and minimum, and median of absolute values of HR, HRV, and filtered ECG.
Additionally, skewness, kurtosis, entropy, and AUC$^{2}$, IQR, pNN20, and pNN50 were extracted. Frequency domain features were also extracted using Welch's method for power spectral analysis and included peak frequencies, absolute powers, normalized powers, and the LF/HF ratio of ultra-low frequency (ULF) band ($< 0.003 Hz$), very low frequency (VLF) band (0.003 Hz - 0.04 Hz), low frequency (LF) band (0.04 Hz - 0.15 Hz), and high frequency (HF) band (0.15 Hz - 0.4 Hz), 
as well as the total power over all frequency bands.

\textbf{EDA Features:} As mentioned in the previous section, the filtered EDA signal is decomposed into a phasic and a tonic responses \cite{schmidt2018introducing, choi2011development, healey2005detecting}. Statistical features, namely, minimum, maximum, mean, median, standard deviation, skewness, kurtosis, entropy, interquartile range, the squared area under the curve, median absolute deviation from EDA signal, phasic, and tonic responses, were calculated. Further, statistical features from the amplitude (excluding the tonic component), height (including the tonic component), recovery time, rise time, and the number of peaks in the phasic component are also computed. 

\textbf{EEG Features:} From four channels of EEG, both time and frequency domain features were computed. For each channel, features from frequency bands: Delta (0.4 - 4) \textit{Hz}, Theta (4 - 8) \textit{Hz}, Alpha (8 - 12) \textit{Hz}, Beta (12 - 31) \textit{Hz}, and Gamma (31 - 128) \textit{Hz} are considered. 
PSD was performed on each frequency band using Welch's method, and each frequency band's absolute, mean, maximum, minimum, and median power was computed. We computed spectral entropy, Hjorth mobility and complexity, Lempel-Ziv Complexity, and Higuchi fractal dimension for the time domain features for each frequency band. 

\textbf{Gaze Features:}
User's gaze features such as pupillary response, blinks, eye fixation, and saccades are correlated with the cognitive load \cite{kramer2020physiological, zagermann2016measuring, chen2011eye}. Therefore, we computed statistical features from left and right pupil diameter, blinks, fixation, and saccades. Further, we computed statistical features from saccade amplitude, velocity, acceleration, de-acceleration, and direction. We also computed the total number of blinks, fixations, and saccades.

\subsection{Dataset Release}
We make the dataset public at:

\href{https://github.com/Prithila05/CLARE}{hhttps://github.com/Prithila05/CLARE}

\section{Data Analysis}
\label{sec:data_analysis}
We examined the subjective cognitive load values recorded from each participant and compared the reported values with the complexity of the segments to better understand the distribution of the reported cognitive load scores. Figure \ref{fig:labeldist} shows the distribution of the cognitive load scores reported for each individual (all sessions are combined). We see that the participants displayed a wide range of distributions, suggesting that the MATB-II software had diverse effects on the cognitive load that the subjects felt. For example, some participants' cognitive load ratings were on the middle or higher end of the PAAS scale (e.g., 1, 2, 3, 14). At the same time, some participants (such as 10, 17, and 19) tended to report numbers on the lower end of the spectrum. We also observed that the participants rarely reported extreme values on the scale.

\begin{figure}
    \centering
        \includegraphics[width=1.1\linewidth]{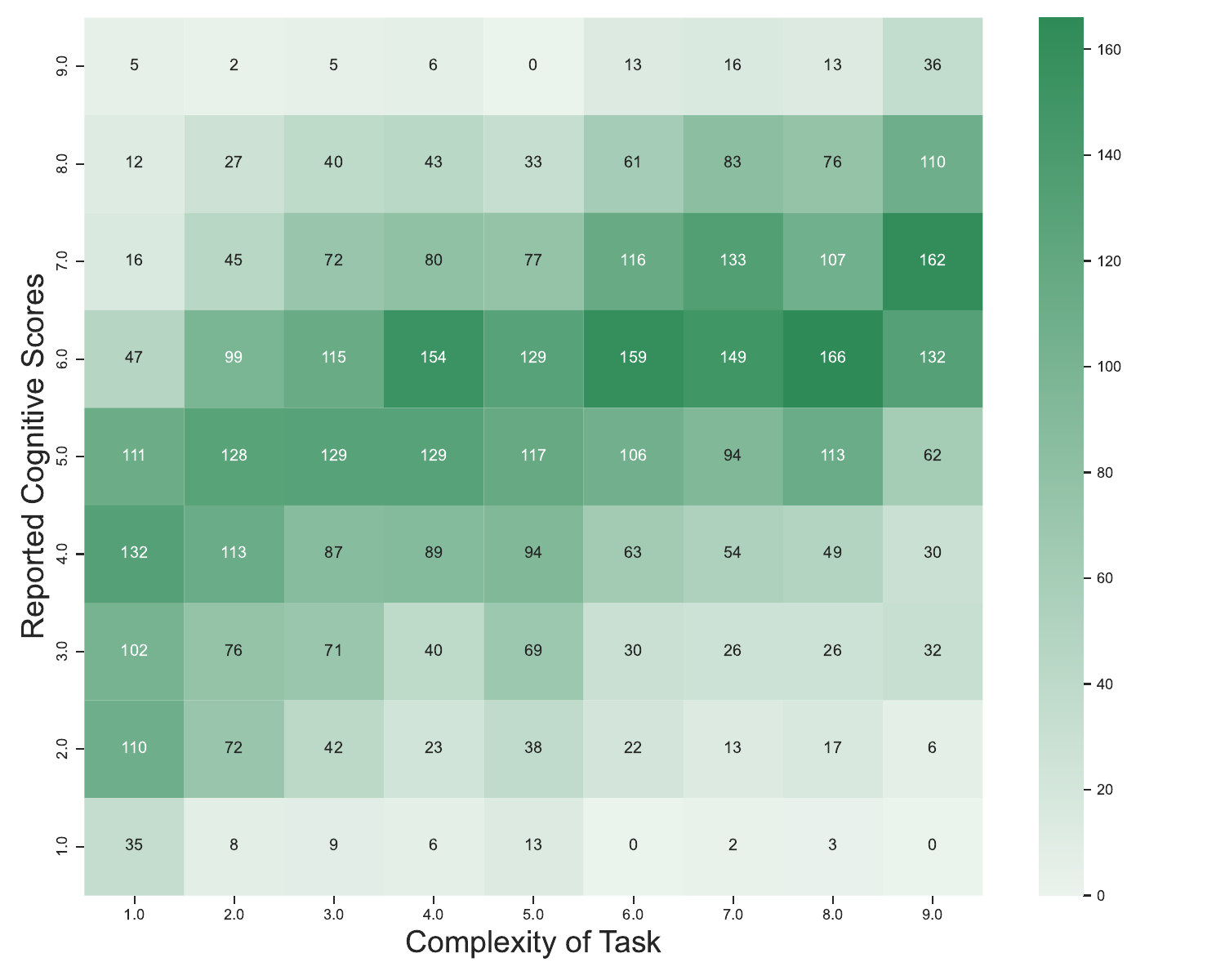}
    \caption{Complexity vs. cognitive load scores. The darker shades mean more samples.}
    \label{fig:complexity_vs_labels}
\end{figure}

We plot a heat map of complexity versus reported cognitive load scores to understand the relation between the complexity of each segment and the associated reported cognitive load values. Figure \ref{fig:complexity_vs_labels} shows the results of this analysis in which we observe an `inverted S` relationship. From the heat map, we can see that for low complexity, most of the cognitive load values reported by participants are concentrated around 1-5; however, as the complexity of the task increases, the cognitive load values reported loosely follow the complexity values. It can also be observed from the map that for the average or higher complexity of tasks, the reported cognitive load values are concentrated between a range of 3 to 7, indicating that most participants experience moderate amounts of cognitive load when complexity is average or higher.

\section{Benchmark Evaluation} \label{sec:results_eval}

\subsection{Classification Algorithms}
We use eight classical machine learning algorithms and two deep learning models for benchmarking. The features extracted from the four modalities (described in Section \ref{subsec:feature_extraction}) serve as inputs to all the models.  
The eight classical machine learning algorithms are Gradient Boosting (GB), Light Gradient Boosting Machine (LGBM), Linear Discriminant Analysis (LDA), Logistic Regression (LR), Multilayer Perceptron (MLP), Random Forest (RF), Support Vector Machine (SVM), and Extreme Gradient Boosting (XGBoost). The deep learning models are a VGG-style convolutional neural network (CNN) \textcolor{black}{and a Transformer network. The selected classical machine learning models have shown strong performances in the past for affective computing from time-series \cite{lima2024multimodal, cheng2017novel}, and each contain different strengths and weaknesses \cite{soofi2017classification}. Similarly, the CNN and Transformer models used are popular deep learning techniques that have been successfully used for affective computing with wearable signals in the past \cite{wu2023transformer, gong2023eeg, wang2023self}, making them reliable choices for benchmarking our dataset's performance. In particular, the selected architectures for the CNN and Transformer have been widely used in prior works \cite{wu2023transformer, gong2023eeg, wang2023self}.}

For the GB classifier, the number of base estimators is set to 300, the loss is set to logarithmic loss, and a maximum depth of 3 is used for individual estimators. For the LGBM, the number of estimators is set to 2000, and for each base estimator, the number of leaves was set to 100. A learning rate of 0.001 is used. The `least squares solution` is selected based on hyperparameter tuning for the LDA classifier. For the LR classifier, the maximum number of iterations required for the solvers to converge is set to 400, and the inverse regularization parameter `C' is set to 1. The MLP classifier is trained with two hidden layers of sizes 100 and 10, and an adaptive learning rate for 1000 iterations. The number of estimators for the RF classifier was set to 1000 with a minimum sample split of 5 and a maximum depth of 5 for each tree. The SVM classifier was trained with the inverse regularization parameter `C' set to 10. For the XGBoost classifier, we used a tree-based boosting method with 300 estimators. We set the learning rate and the L1 regularization values to 1e-3 and 1e-4, respectively. The models' parameters were tuned empirically to obtain the best training results.

\begin{figure}
    \centering
    \includegraphics[width=1\linewidth]{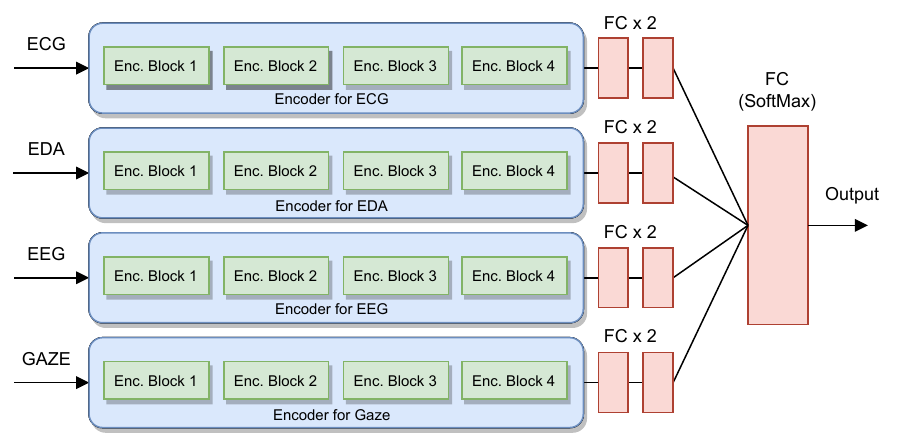}
    \caption{A multimodal pipeline consisting of four Encoders, one for each modality.}
    \label{fig:arch}
\end{figure}

\textcolor{black}{
The CNN consists of 4 convolution blocks, each containing two 1D convolutional layers with a ReLU activation layer and a MaxPool layer of filter size and stride of 2. The output of the fourth encoder block is fed to two FC layers with 512 and 256 units, respectively. A feature-level fusion strategy (late fusion) is then used to create the multimodal setup. The final output class is generated by SoftMax following an FC layer.}
Figure \ref{fig:arch} shows the multimodal pipeline with the four recorded modalities. The details of filters, filter sizes, and strides for each encoder block are presented in Table \ref{tbl:archdetails}. Similar to the machine learning models, the hyperparameters of the CNN were empirically tuned to achieve the best performance.

\textcolor{black}{The Transformer network \cite{vaswani2017attention} consists of four Transformer blocks and a prediction head. The output from the Transformer blocks is flattened and fed into the prediction head. The prediction head contains two prediction blocks consisting of FC layers of size 256 and 128 followed by ReLU and dropout of 0.5. The output layer contains a sigmoid function. For the multimodal setup, all the features from the individual modalities are concatenated before feeding the Transformer network (early fusion). We used an early fusion setup here as it yielded better results in comparison to the late fusion setup used for the CNN. The entire sequence is then considered as one patch in order to capture the global context of the entire sequence and processed all together by the Transformer network.}

\begin{figure}
    \centering
    \includegraphics[width=1\linewidth]{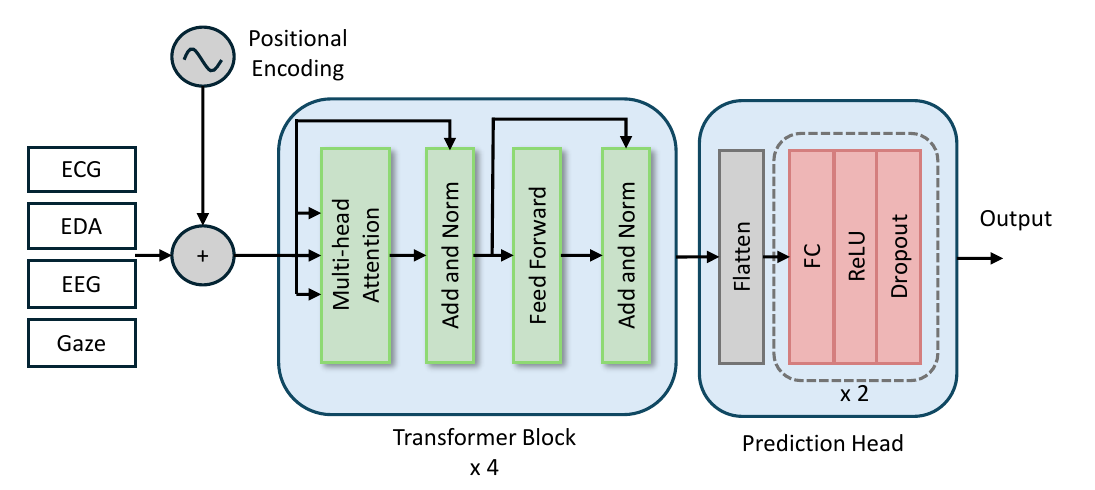}
    \caption{Multimodal pipeline of the transformer network.}
    \label{fig:transformer}
\end{figure}

\subsection{Evaluation Protocol and Implementation Details}
\textcolor{black}{For all the benchmarks, we use binary (`high' versus `low') cognitive load classification. To calculate binary labels, reported scores of less than 5 were considered `low' cognitive load, and scores of 5 or more were considered `high' cognitive load. This threshold is chosen because it provides a clear and straightforward division between lower and higher levels of cognitive load, allowing us to perform binary classification. Our post-collection interviews with participants indicated that many participants attributed 5 with a relatively high cognitive load state, and therefore we placed 5 and higher in the `high' category, while 1 to 4 were placed in the `low' category.
}
We use 10-fold and Leave-One-Subject-Out (LOSO) cross-validation evaluation schemes to indicate the performance of our models on previously unseen data. 
Accuracy and F1 score were used as evaluation metrics. 
For all the machine learning algorithms except LGBM and XGBoost, we use the Scikit-learn\footnote{https://scikit-learn.org/stable/} library. All the parameters of the machine learning algorithms were optimized by performing an exhaustive search.
Our deep learning pipelines were implemented using the TensorFlow framework. For the CNN, we used AdaDelta with a decay rate of 0.95, a learning rate of 5e-3, and focal loss (alpha of 4.0 and gamma of 2.0). 
\textcolor{black}{For the transformer network we used Adam \cite{kingma2014adam} optimizer with a learning rate of 0.0001.}
We empirically selected a batch size of 256 for 100 epochs for all experiments. We implemented our pipelines using an Intel Core i7-9700 CPU and an NVIDIA GeForce RTX 2080 Ti GPU.

\begin{table*}
	\caption{Classifiers performance for binary classification of high and low cognitive load using 10-fold evaluation scheme, the numbers are in accuracy (F1) format.}
	\scriptsize
        \centering
\setlength{\tabcolsep}{3pt}        
	\begin{tabular}{lcccccccccc}
	\midrule[.0025\linewidth]
		\multirow{2}{*}{\textbf{Modalities}} & \multicolumn{9}{c}{\textbf{Models}} \\ \cmidrule{2-11}
		 & GB & LGBM	& LDA &	LR & MLP & RF & SVM & XGBoost & CNN & \textcolor{black}{Transformer}\\
		\midrule[.0025\linewidth]
		ECG & 66.58 (65.49) & 	66.71 (66.27) & 	60.49 (57.62) & 	60.73 (58.21) & 	62.19 (61.07) & 	68.28 (67.38) & 	64.20 (63.83) & 	66.68 (65.26) & \textbf{78.45 (76.41)} & \textcolor{black}{\underline{77.40(66.32)}}\\
		EDA & 68.34 (67.39) & 	70.18 (69.64) & 	63.52 (61.65) & 	63.48 (61.53) & 	60.56 (59.03) &  \underline{71.98 (70.93)} & 	65.59 (65.00) & 	69.33 (68.06) & 68.01 (66.21) & \textcolor{black}{\textbf{74.05(52.84)}}\\
		EEG & 69.19 (68.36) & 	69.94 (69.42) & 	64.30 (63.02) & 	64.84 (63.62) & 	62.53 (61.94) & 	71.85 (71.02) & 	69.09 (68.67) & 	69.91 (69.00) & \underline{74.10 (72.08)} & \textcolor{black}{\textbf{77.37(71.84)}}\\
		Gaze & 66.85 (66.01) & 	67.97 (67.45) & 	62.39 (60.11) & 	62.50 (60.35) & 	61.54 (59.61) & 	\underline{69.70 (68.66)} & 	65.83 (65.51) & 	67.80 (66.72) & 66.78 (64.23) &  \textcolor{black}{\textbf{74.66(64.88)}}\\
		\midrule
		ECG, EDA & 70.08 (69.31) & 	71.98 (71.56) & 	64.37 (62.89) & 	64.67 (63.20) & 	59.95 (58.15) & 	73.24 (72.51) & 	69.30 (68.83) & 	70.96 (70.15) & \underline{79.72 (77.53)} & \textcolor{black}{\textbf{80.85(73.55)}}\\
		ECG, EEG & 70.76 (69.99) & 	70.18 (69.67) & 	65.66 (64.59) & 	65.93 (64.96) & 	63.38 (62.56) & 	72.53 (71.77) & 	68.75 (68.33) & 	69.30 (68.32)  & \underline{77.66 (75.77)} & \textcolor{black}{\textbf{81.85(75.60)}}\\
		ECG, Gaze & 68.96 (68.11) & 	70.38 (69.92) & 	63.96 (62.71) & 	64.13 (62.92) & 	63.86 (62.67) & 	71.00 (70.06) & 	67.80 (67.33) & 	70.59 (69.71) & \underline{78.41 (76.54)} & \textcolor{black}{\textbf{81.00(74.50)}}\\
		EDA, EEG & 72.19 (71.39) & 	72.09 (71.63) & 	65.35 (64.21) & 	65.28 (64.22) & 	60.96 (59.17) & 	73.61 (72.85) & 	68.86 (68.40) & 	72.49 (71.65) & \underline{75.30 (73.45)} & \textcolor{black}{\textbf{82.41(76.76)}}\\
		EDA, Gaze & 71.34 (70.63) & 	72.32 (71.87) & 	65.45 (64.12) & 	65.73 (64.43) & 	57.67 (54.13) & 	\underline{74.40 (73.65)} & 	68.85 (68.34) & 	72.63 (71.83) & 70.64 (68.23) & \textcolor{black}{\textbf{81.85(74.99)}}\\
		EEG, Gaze & 71.64 (70.92) & 	71.71 (71.21) & 	66.98 (66.08) & 	67.43 (66.48) & 	66.30 (64.85) & 	73.65 (72.76) & 	70.79 (70.40) & 	71.40 (70.59) & \underline{75.50 (73.13)} & \textcolor{black}{\textbf{82.59(76.79)}}\\
        \midrule
		ECG, EDA, EEG &  71.10 (70.42) & 	72.90 (72.57) & 	66.13 (65.07) & 	65.73 (64.74) & 	61.71 (60.51) & 	73.65 (72.97) & 	69.87 (69.43) & 	72.53 (71.72) & \underline{80.10 (78.49)} & \textcolor{black}{\textbf{84.12(80.08)}}\\
		ECG, EDA, Gaze & 72.12 (71.45) & 	73.24 (72.86) & 	67.12 (66.04) & 	66.61 (65.60) & 	62.39 (60.10) & 	74.53 (73.85) & 	70.96 (70.49) & 	73.27 (72.54) & \underline{80.30 (78.34)} & \textcolor{black}{\textbf{83.43(78.66)}}\\
		ECG, EEG, Gaze & 71.51 (70.82) & 	72.80 (72.34) & 	67.56 (66.74) & 	68.51 (67.67) & 	66.54 (65.14) & 	74.09 (73.26) & 	71.13 (70.80) & 	71.64 (70.85) & \underline{78.26 (76.56)} & \textcolor{black}{\textbf{83.63(78.39)}}\\
		EDA, EEG, Gaze & 73.58 (72.96) & 	73.58 (73.13) & 	67.97 (67.13) & 	67.87 (67.03) & 	61.51 (59.51) & 	75.18 (74.45) & 	71.71 (71.26) & 	74.50 (73.78) & \underline{75.87 (74.08)} & \textcolor{black}{\textbf{84.93(78.18)}}\\
        \midrule
		All modalities & 73.38 (72.81) & 	74.26 (73.83) & 	68.24 (67.47) & 	68.38 (67.59) & 	62.36 (60.66) & 	75.14 (74.47) & 	72.05 (71.59) & 	74.26 (73.60) & \underline{79.65 (77.94)} & \textcolor{black}{\textbf{85.58(81.18)}}\\
		\bottomrule[.0025\linewidth]
	\end{tabular}
	\label{tbl:kfold}
\end{table*}

\begin{table*}
	\caption{Classifiers performance for binary classification of high and low cognitive load using leave-one-subject-out evaluation scheme, the numbers are in accuracy (F1) format.}
	\scriptsize
\setlength{\tabcolsep}{3pt}        
        \centering
	\begin{tabular}{lcccccccccc}
	\midrule[.0025\linewidth]
		\multirow{2}{*}{\textbf{Modalities}} & \multicolumn{9}{c}{\textbf{Models}} \\ \cmidrule{2-11}
		 & GB & LGBM	& LDA &	LR & MLP & RF & SVM & XGBoost & CNN & \textcolor{black}{Transformer}\\
		\midrule[.0025\linewidth]
		ECG & 53.23 (45.21) & 42.00 (33.06) & 55.60 (43.79) & 43.34 (35.22) & 43.15 (39.08) & 55.60 (44.16) & 40.26 (33.09) & 40.00 (28.79) & \underline{63.28 (51.00)} & \textcolor{black}{\textbf{66.15(54.20)}}\\
		EDA & 58.67 (50.86) & 	55.20 (51.03) & 	60.15 (51.23) & 	50.81 (47.56) & 	48.90 (43.91) & 	59.36 (49.50) & 	50.46 (47.42) & 	51.52 (47.67) & \textbf{64.08 (54.80)} & \textcolor{black}{\underline{63.73(54.43)}}\\
		EEG & 54.71 (47.06) & 	53.35 (46.40) & 	49.76 (43.83) & 	52.38 (44.16) & 	54.46 (48.94) & 	59.17 (46.57) & 	53.23 (47.03) & 	56.54 (48.95) & \underline{65.21 (52.15)} & \textcolor{black}{\textbf{67.77(57.03)}}\\
		Gaze & 54.61 (47.35) & 54.79 (46.41) & 56.72 (42.49) & 52.79 (43.97) & 53.08 (44.76) & 58.80 (48.83) & 51.63 (44.05) & 56.93 (46.53) & \underline{60.23 (52.92)} & \textcolor{black}{\textbf{65.87(51.73)}}\\
		\midrule
		ECG, EDA & 58.66 (50.54) & 	45.88 (38.04) & 58.37 (47.25) & 51.87 (48.68) & 47.60 (44.53) & 	62.23 (51.60) & 50.58 (47.58) & 47.50 (43.31) & \underline{64.48 (55.38)} & \textcolor{black}{\textbf{68.99(62.32)}}\\
		ECG, EEG & 55.10 (48.49) & 	53.54 (46.77) & 	50.27 (44.37) & 	51.97 (44.11) & 	52.17 (47.11) & 	56.32 (43.89) & 	49.29 (43.36) & 56.97 (50.15) & \underline{65.22 (51.00) } & \textcolor{black}{\textbf{68.74(60.72)}}\\
		ECG, Gaze & 54.29 (49.07) & 	56.62 (50.75) & 	57.21 (47.22) & 	57.22 (47.42) & 	53.65 (48.40) & 	57.66 (47.21) & 	54.51 (47.36) & 	56.59 (49.79) & \underline{59.27 (50.81)} & \textcolor{black}{\textbf{68.02(59.82)}}\\
		EDA, EEG & 56.53 (49.60) & 	57.91 (52.19) & 	51.51 (44.89) & 	51.27 (44.51) & 	52.80 (46.45) & 	60.33 (48.58) & 	53.19 (47.38) & 	62.30 (54.40) & \underline{65.94 (56.46)} & \textcolor{black}{\textbf{70.45(65.00)}}\\
		EDA, Gaze & 59.74 (54.60) & 	60.80 (53.98) & 	58.28 (49.39) & 	57.55 (48.64) & 	54.12 (47.70) & 	63.88 (53.86) & 	56.23 (49.43) & 	62.29 (55.87) & \underline{63.95 (56.89)} & \textcolor{black}{\textbf{69.59(63.20)}}\\
		EEG, Gaze & 60.63 (53.31) & 	60.01 (53.18) & 	57.74 (48.44) & 	57.57 (49.15) & 	58.90 (49.53) & 	\underline{61.41 (48.68)} & 	57.50 (49.98) & 	60.22 (52.86) & 60.90 (52.04) & \textcolor{black}{\textbf{69.22(59.51)}}\\
		\midrule
		ECG, EDA, EEG & 58.34 (51.06) & 	59.96 (52.88) & 	52.00 (45.33) & 	50.72 (44.10) & 	53.38 (47.55) & 	60.54 (48.45) & 	53.48 (46.50) & 	61.66 (54.38) & \underline{65.48 (55.72)} & \textcolor{black}{\textbf{70.90(66.84)}}\\
		ECG, EDA, Gaze & 59.79 (52.34) & 	60.47 (54.12) & 	57.20 (47.83) & 	56.51 (47.48) & 	58.71 (49.91) & 	\underline{63.02 (51.48)} & 	58.05 (51.30) & 	61.04 (54.06) & 62.08 (55.10) & \textcolor{black}{\textbf{70.61(65.91)}}\\
		ECG, EEG, Gaze & 58.54 (51.16) & 	58.37 (52.30) & 	57.63 (49.93) & 	57.74 (50.91) & 	56.78 (48.50) & 	60.34 (47.60) & 	54.73 (47.46) & 	59.99 (53.31) & \underline{60.97 (51.51)} & \textcolor{black}{\textbf{69.27(62.20)}}\\
		EDA, EEG, Gaze & 61.90 (54.95) & 	61.55 (55.55) & 	56.63 (48.09) & 	56.38 (48.92) & 	52.31 (45.70) & 	62.25 (50.91) & 	58.42 (50.96) & 	61.95 (55.68) & \underline{63.65 (56.13)} & \textcolor{black}{\textbf{72.15(68.50)}}\\
		\midrule		
		All modalities & 57.95 (51.58) & \underline{63.91 (56.59) }& 56.35 (50.52) & 54.72 (49.67) & 54.89 (48.94) & 62.73 (51.58) & 58.63 (50.49) & 63.75 (56.95) & 61.38 (55.34) & \textcolor{black}{\textbf{72.70(69.46)}}\\
		\bottomrule[.0025\linewidth]
	\end{tabular}
	\label{tbl:loso}
\end{table*}

\subsection{Results}
This section provides the uni-modal, and multimodal benchmark binary classification results from machine learning and the deep learning classifiers with two evaluation schemes, namely 10-fold and the LOSO evaluation scheme. The benchmark results for 10-fold and LOSO schemes are provided in Table \ref{tbl:kfold} and \ref{tbl:loso}, respectively. Below we discuss these results in detail.

\subsubsection{10-Fold Cross Validation} 
\textcolor{black}{In this evaluation scheme, we observe that the Transformer generally has the best performance with an accuracy of 85.58 and an F1 of 81.18 using all the modalities, as well as the most consistent performance for both accuracy and F1. This is followed by the CNN which achieves an accuracy of 80.30 and F1 of 78.34 with ECG, EDA, and Gaze. The CNN, outperforms the Transformer when using ECG. In terms of individual modalities, it can be seen the ECG is the most important, with accuracy and F1 scores of 78.45 and 76.41, respectively, for the CNN. In the bi-modal setup, we observe that EEG and Gaze show the highest results with an accuracy of 82.59 and an F1 of 76.79. This is followed by EDA and EEG which obtain an accuracy of 82.41 and an F1 of 76.76. The highest results for tri-modal learning are achieved with ECG, EDA, and EEG (accuracy = 84.12, F1 = 80.08), closely followed by EDA, EEG, and Gaze (accuracy = 84.93, F1 = 78.18).}

\subsubsection{LOSO Cross Validation}
\textcolor{black}{
We observe that in the LOSO setup, the deep learning methods show higher improvements with respect to the classical machine learning method, compared to the 10-fold scheme. 
Consistent with our observations for 10-fold evaluation, we observe that for LOSO, the Transformer has the best performance with an accuracy of 72.70 and an F1 of 69.46, followed by the CNN and RF methods. In one case (EDA) the CNN classifier outperforms the Transformer. In some cases, the RF classifier (EEG \& Gaze, and ECG, EDA \& Gaze) and the LGBM classifier (all modalities) outperform the CNN. In terms of individual modalities, we observe that EEG is the most important modality and yields the best results with accuracy and F1 scores of 67.77 and 57.03, respectively. Following the same trend, for bi-modal learning, the best results are achieved using EDA and EEG (accuracy = 70.45, F1 = 65.00) followed by EDA and Gaze (accuracy = 69.59, F1 = 63.20). The best results for tri-modal learning are achieved with EDA, EEG and Gaze (accuracy = 72.15, F1 = 68.50).}

\subsection{Inference Time}
As mentioned earlier, in this experiment, the ground truth labels were recorded at 10-second intervals to facilitate the collection of the frequent ground truth labels. This allows for better estimation of cognitive load scores by machine learning models. To meet the real-time inference requirements, the model constraint would be to generate the output prediction in less time than the ground truth collection interval. For classical machine learning models, the inference time would depend on the time to compute handcrafted features and the time to make a prediction. In contrast, for the deep learning model, as there is no computation of handcrafted features, the inference time would depend on the time taken to make a prediction. So, we extract low complexity features for classical machine learning models \cite{loconsole2014real}. Given the number of handcrafted features from four modalities, the average time to compute these features for one sample was approximately 1 second. We then evaluated and present the average inference time of each classical machine learning model and deep learning model in Table \ref{tbl:inference}. 
\textcolor{black}{The table shows that RF, XGBoost, CNN and Transformer models take significantly more time than the other methods; however, the inference time is still under 1 second for all the algorithms that satisfy the real-time inference requirements given that each segment is 10 seconds long.}

{\renewcommand{\arraystretch}{1.2}
\begin{table} 
\begin{center}
\caption{Average Inference Time}

\scriptsize
\begin{tabular}{cc} 
\midrule[.15em]
\multirow{1}{*}{\textbf{Classifiers}} & \multicolumn{1}{c}{\textbf{Inference Time} ($\mu$s)} \\ \midrule[.001\linewidth]
Gradient Boosting & 20.99\\
Light Gradient Boosting Machine & 64.17\\
Linear Discriminant Analysis & 15.97\\
Logistic Regression & 16.47\\
Multi-Layer Perceptron & 20.52\\
Random Forest & 605.88\\
Support Vector Machine  & 446.64\\
Extreme Gradient Boosting & 1482.30\\
CNN & 4519.99\\
\textcolor{black}{Transformer} & \textcolor{black}{4579.37} \\
\bottomrule[.001\linewidth]
\end{tabular}
\label{tbl:inference}
\end{center}
\end{table}}

\section{Conclusion and Future Directions} \label{sec:conclusion}
In summary, we collected a new multimodal dataset, CLARE, with self-reported \textit{cognitive load} as ground-truth labels. The dataset comprised four modalities (ECG, EDA, EEG, and Gaze) from 24 participants. We used MATB-II software to develop nine different complexity levels, each with a duration of 1 minute, to induce varying levels of mental load on participants. Each participant completed four separate sessions comprising nine randomly selected 1-minute segments. They reported their subjective cognitive load every 10 seconds during the experiment as physiological signals were recorded. To evaluate our dataset, we provide uni-modal and multimodal benchmarks from machine learning and deep learning classifiers with a 10-fold and leave-one-subject-out evaluation scheme.

We identify a few limitations and future directions for the dataset collection and design. As explained earlier, participants were asked to report their cognitive load score at 10-second intervals during the experiments. While this design intended for frequent ground-truth labels to be recorded, which would benefit machine learning models, it may have induced an additional unwanted cognitive load on the participants. For future similar data collection protocols, studies can be done to identify the optimum balance between the frequency of collected labels and the added distraction caused by the collection effort. Such a study can inform more optimized protocols for cognitive load and affective computing studies.

The experiment results suggest that the accurate, real time evaluation and presentation of cognitive load is possible and that the outputs can potentially be used to support cognitive performance in a broad range of contexts. The results also suggest that in terms of number of signals/modalities ``more is not always better'' and accurately indexing cognitive load can be accomplished with a carefully selected subset of physiological signals.

\section{Acknowledgement}
We would like to thank the Innovation for Defence Excellence and Security (IDEaS) program under the Department of National Defence (DND) for funding this project.

\appendices
\appendices

\bibliographystyle{IEEEtran}
\bibliography{ref}

\end{document}